\documentclass[lettersize,journal,comsoc]{IEEEtran}
\usepackage{amsmath,amsfonts}
\usepackage{array}
\usepackage[caption=false,font=normalsize,labelfont=sf,textfont=sf]{subfig}

\usepackage{booktabs}
\usepackage{makecell}

\usepackage[english]{babel}
\usepackage[noadjust]{cite}
\usepackage[linesnumbered,ruled,vlined]{algorithm2e}
\usepackage{algpseudocode}
\usepackage{optidef}

\ifCLASSINFOpdf
\else
   \usepackage[dvips]{graphicx}
\fi
\usepackage{url}

\usepackage{multirow}
\usepackage{tabularx}
\usepackage{optidef}
\usepackage{amsmath}
\usepackage{bm}
\usepackage{amssymb}
\usepackage{xcolor}
\usepackage{caption}
\usepackage{subcaption}
\usepackage{graphicx}

\usepackage{lipsum}
\usepackage{amsthm}
\usepackage{siunitx}
\usepackage[acronym]{glossaries}

\usepackage[hidelinks]{hyperref}
\usepackage[acronym,automake]{glossaries-extra}

\usepackage{titlesec}
\titlespacing\section{0pt}{*0.7}{*0.7}
\titlespacing\subsection{0pt}{*0.7}{*0.7}
\titlespacing\subsubsection{0pt}{*0.6}{*0.6}

\theoremstyle{definition}

\newtheorem{lemma}{Lemma}

\makeatletter
\newrobustcmd*{\myGls}{\@gls@hyp@opt\@myGls}
\newcommand*{\@myGls}[2][]{%
  \new@ifnextchar[{\@myGls@{#1}{#2}}{\@myGls@{#1}{#2}[]}%
}
\def\@myGls@#1#2[#3]{%
  \glsdoifexists{#2}%
  {%
    \let\do@gls@link@checkfirsthyper\@gls@link@checkfirsthyper
    \let\glsifplural\@secondoftwo
    \let\glscapscase\@secondofthree
    \def\glscustomtext{%
    \ifglsused{#2}
    {\acronymfont\glsentryshort{#2}#3}
    {%
    \ecapitalisewords{\glsentrylong{#2}}#3\space%
    \firstacronymfont(\glsentryshort{#2})}%
    }%
    \def\glsinsert{#3}%
    \def\@glo@text{\csname gls@\glstype @entryfmt\endcsname}%
    \@gls@link[#1]{#2}{\@glo@text}%
    \ifKV@glslink@local
      \glslocalunset{#2}%
    \else
      \glsunset{#2}%
    \fi
  }%
  \glspostlinkhook
}

\newrobustcmd*{\myGlspl}{\@gls@hyp@opt\@myGlspl}
\newcommand*{\@myGlspl}[2][]{%
  \new@ifnextchar[{\@myGlspl@{#1}{#2}}{\@myGlspl@{#1}{#2}[]}%
}
\def\@myGlspl@#1#2[#3]{%
  \glsdoifexists{#2}%
  {%
    \let\do@gls@link@checkfirsthyper\@gls@link@checkfirsthyper
    \let\glsifplural\@firstoftwo
    \let\glscapscase\@secondofthree
    \def\glscustomtext{%
    \ifglsused{#2}%
    {\acronymfont\glsentryshortpl{#2}#3}
    {%
    \ecapitalisewords{\glsentrylongpl{#2}}#3\space%
    \firstacronymfont(\glsentryshortpl{#2})}%
    }%
    \def\glsinsert{#3}%
    \def\@glo@text{\csname gls@\glstype @entryfmt\endcsname}%
    \@gls@link[#1]{#2}{\@glo@text}%
    \ifKV@glslink@local
      \glslocalunset{#2}%
    \else
      \glsunset{#2}%
    \fi
  }%
  \glspostlinkhook
}
\makeatother

\setabbreviationstyle[acronym]{long-short}

\makeglossaries

\newacronym{aoa}{AoA}{angle-of-arrival}
\newacronym{aod}{AoD}{angle-of-departure}
\newacronym{bs}{BS}{base station}
\newacronym{psr}{PSR}{power split ratio}
\newacronym{isac}{ISAC}{integrated sensing and communications}
\newacronym{mimo}{MIMO}{multiple-input multiple-output}
\newacronym{iv}{IV}{instrumental variable}
\newacronym{nsp}{NSP}{null-space projection}
\newacronym{dof}{DoF}{degree of freedom}
\newacronym{sinr}{SINR}{signal-to-interference-plus-noise ratio}
\newacronym{sdma}{SDMA}{spatial division multiplexing}
\newacronym{csi}{CSI}{channel state information}
\newacronym{mui}{MUI}{multi-user interference}
\newacronym{rcg}{RCG}{Riemannian conjugate gradient}
\newacronym{crb}{CRB}{Cramer-Rao bound}
\newacronym{lmmse}{LMMSE}{linear minimum mean square error}
\newacronym{isl}{ISL}{integrated sidelobe level}
\newacronym{papr}{PAPR}{peak-to-average power ratio}
\newacronym{swipt}{SWIPT}{simultaneous information and power transfer}
\newacronym{qos}{QoS}{quality-of-service}
\newacronym{sar}{SAR}{synthetic aperture radar}
\newacronym{svd}{SVD}{singular value decomposition}
\newacronym{xr}{XR}{extended reality}
\newacronym{3gpp}{3GPP}{Third Generation Partnership Project}
\newacronym{nr}{NR}{New Radio}
\newacronym{prs}{PRS}{Positioning Reference Signal}
\newacronym{etsi}{ETSI}{European Telecommunication Standard Institution}
\newacronym{kkt}{KKT}{Karush-Kuhn-Tucker}
\newacronym{mm}{MM}{majorization-minimization}
\newacronym{mse}{MSE}{mean square error}
\newacronym{sdr}{SDR}{semi-definite relaxation}
\newacronym{itu}{ITU}{International Telecommunication Union}
\newacronym{mmse}{MMSE}{minimum mean square error}
\newacronym{nmse}{NMSE}{normalized mean square error}
\newacronym{xpd}{XPD}{cross-polarization discrimination}
\newacronym{snr}{SNR}{signal-to-noise ratio}
\newacronym{lmi}{LMI}{linear matrix inequality}
\newacronym{qcqp}{QCQP}{quadratically constrained quadratic programming}
\newacronym{sdp}{SDP}{semidefinite programming}
\newacronym{capex}{CAPEX}{capital expenditure}
\newacronym{ris}{RIS}{reconfigurable intelligent surface}
\newacronym{ipsac}{IPSAC}{integrated polarimetric sensing and communication}
\newacronym{ula}{ULA}{uniform linear array}
\newacronym{upa}{UPA}{uniform planar array}

\newcommand{\bb}[1]{\textcolor{black}{#1}}

\allowdisplaybreaks[4]
\begin{document}
\title{Integrated Polarimetric Sensing and Communication with Polarization-Reconfigurable Arrays}

\author{
Byunghyun Lee, Rang Liu, David J. Love, James V. Krogmeier, and A. Lee Swindlehurst
\thanks{
This work is supported in part by the National Science Foundation under grants EEC-1941529 and CNS-2212565.
}
\thanks{Byunghyun Lee was with the Elmore Family School of Electrical and Computer
Engineering, Purdue University, West Lafayette, IN 47907 USA. He is now with Qualcomm Technologies, Inc.
5775 Morehouse Drive, San Diego, CA 92121 (e-mail: byunlee@qti.qualcomm.com).}
\thanks{David J. Love and James V. Krogmeier are with the Elmore Family School of Electrical and Computer Engineering, Purdue University, West Lafayette, IN 47907 USA (e-mails:
\{ djlove, jvk\}@purdue.edu).
}
\thanks{R. Liu and A. Lee Swindlehurst are with the Center for Pervasive Communications and Computing, University of California, Irvine, CA 92697, USA (e-mails: \{rangl2, swindle\}@uci.edu).}
}

\maketitle

\begin{abstract}

Polarization diversity offers a cost- and space-efficient solution to enhance the performance of integrated sensing and communication systems.
Polarimetric sensing exploits the signal's polarity to extract details about the target such as shape, pose, and material composition.
From a communication perspective, polarization diversity can enhance the reliability and throughput of communication channels.
This paper proposes an \gls{ipsac} system that jointly conducts polarimetric sensing and communications.
We study the use of single-port polarization-reconfigurable antennas to adapt to channel depolarization effects, without the need for separate RF chains for each polarization.
We address two core sensing tasks in \gls{ipsac} systems, target parameter estimation and target detection.
For parameter estimation, we consider the problem of minimizing the \gls{mse} of the target depolarization parameter estimate, which is a critical task for various polarimetric radar applications such as rainfall forecasting, vegetation identification, and target classification.
To address this nonconvex problem, we apply \gls{sdr} and \gls{mm} optimization techniques.
Next, we consider a design that maximizes the target \gls{sinr} leveraging prior knowledge of the target and clutter depolarization statistics to enhance the target detection performance.   
To tackle this problem, we modify the solution developed for \gls{mse} minimization subject to the same \gls{qos} constraints.
Extensive simulations show that the proposed polarization reconfiguration method substantially improves the depolarization parameter \gls{mse}.
Furthermore, the proposed method considerably boosts the target \gls{sinr} due to polarization diversity, particularly in cluttered environments.

\end{abstract}

\begin{IEEEkeywords}
integrated sensing and communications, multiple-input multiple-output, polarization
\end{IEEEkeywords}

\IEEEpeerreviewmaketitle
\glsresetall

\section{Introduction}

\Gls{isac} is a key technology for 6G communication systems, enabling sensing functionality using wireless networks \cite{brinton2024key6g}.
By sharing physical (e.g., antennas) and spectral resources and even signal waveforms, \gls{isac} enhances spectral and energy efficiency.
The additional sensing capabilities offered by ISAC, including detection, localization, and imaging, enable environment-aware, cognitive wireless systems that can support emerging applications such as precision agriculture, robotics, autonomous vehicles, and extended reality.  
Driven by these promising prospects, standardization bodies and industry players have begun incorporating \gls{isac} into communication standards and commercial systems \cite{ITU_IMT_2030_Framework_2023}.





\Gls{mimo} technology is expected to remain fundamental to the evolution of both sensing and communications.
Increasing the number of antennas significantly improves spatial resolution and coverage, benefiting both sensing and communications.
\bb{
In line with this, researchers have actively studied various aspects of \gls{isac} systems such as beamformer design, power allocation, antenna placement optimization, and joint channel and target parameter estimation \cite{lee2024constant,leeSpatialDivisionISACPractical2025a,liuDOAEstimationOrientedJoint2024,liuDualFunctionalRadarCommunicationWaveform2021,nguyenPerformanceAnalysisPower2025,huaSecureIntelligentReflecting2024,zhangIntegratedSensingCommunication2024b}.
}
However, employing a massive number of antennas comes with increased hardware costs (e.g., RF chains), bulkier circuits, complexity, and power consumption, which hinder practical implementation and scalability.

Polarization diversity has been recognized as an effective approach for alleviating limitations on the available spatial degrees of freedom (DoFs) associated with traditional MIMO arrays, providing additional DoFs in the polarization domain without expanding the array size.
Polarization refers to the orientation of the electric field vector relative to the direction of propagation of an electromagnetic wave.
\textit{Polarimetric radar sensing} was first introduced in the 1950s to obtain more precise information about target objects by analyzing the state of polarized transmissions after interaction with the target \cite{hunterPolarizationRadarEchoes1954}.
%
A fundamental parameter in polarimetric radar is the \textit{depolarization matrix}, which describes the depolarization effect of the target and conveys information about the target's shape, orientation, reflectivity, and material composition. 
A wide range of polarimetric radar applications have been studied, such as rainfall forecasting, target classification, material identification, and radar imaging \cite{notarosPolarimetricWeatherRadar2022b,torvikClassificationBirdsUAVs2016,lee2017polarimetric}. 

%


From a communication perspective, dual-polarized \gls{mimo}, which uses colocated antennas with orthogonal polarizations, has been a common solution for commercial systems due to its enhanced transmit/receive diversity and improved channel capacity \cite{nabarPerformanceMultiantennaSignaling2002,oestgesDualpolarizedWirelessCommunications2008,kimLimitedFeedbackBeamforming2010}.
Along this line, the incorporation of dual-polarization signaling in commercial mobile broadband has been extensively studied, including codebook design and channel modeling \cite{3GPP-R1-105011,3gpp.38.901}.
To fully exploit polarization diversity, dual-polarized MIMO requires independent data streams fed separately to each polarization branch.
This requires twice as many RF chains as in single-polarized systems, greatly increasing hardware cost, complexity, and power consumption  \cite{castellanosLinearPolarizationOptimization2024,doanAchievableCapacityMultipolarization2021}.
Consequently, the practical deployment of dual-polarized arrays becomes challenging, particularly in scenarios where massive arrays are employed.


\bb{
To overcome these limitations, new solutions have been explored to achieve polarization adaptability without doubling the RF chains.
One such approach employs movable antennas that physically align their polarization with that of the incoming signal by adjusting their position and orientation \cite{shao2025polarized,
zhang2024polarization}.
However, movable antennas rely on mechanical components, which inherently lead to higher energy consumption and hardware cost compared to conventional antennas.
As an alternative, single-port polarization-reconfigurable arrays have been studied in recent communications research \cite{castellanosLinearPolarizationOptimization2024,doanAchievableCapacityMultipolarization2021,zhou2024polarforming,kwonCapacityMaximizationPolarizationAgile2015,zhuDesignPolarizationReconfigurable2014}.
In such arrays, each antenna can electronically adjust the polarization state by integrating orthogonally polarized elements connected via a single RF chain. 
Therefore, this approach effectively addresses the challenges of traditional dual-polarized \gls{mimo} systems with significantly reduced hardware cost and power consumption.
}

Despite its well-established advantages in radar and communication systems, the role of polarization remains underexplored in the \gls{isac} domain.
\bb{
Several recent ISAC studies have addressed aspects such as dual-polarized channel modeling \cite{lvDualPolarizationChannelModeling2024}, power control \cite{shaoRobustPowerControl2025}, receiver orientation estimation \cite{ibrahimInferringDirectionOrientation2024}
and secrecy sum rate maximization
\cite{zhangDualPolarizedRISAssistedSecure2024}.}
In addition, 
\cite{xia2025ris} considered a separate beam pattern design for vertical and horizontal polarizations to simultaneously detect two different targets.
However, these works primarily exploit orthogonal polarizations without explicitly considering the essential polarimetric sensing characteristics captured by the depolarization matrix. 
Furthermore, polarization-reconfigurable antenna arrays, which have demonstrated substantial potential in communication-only scenarios, have not yet been explored for ISAC systems. 
Therefore, the potential of fully integrated polarimetric sensing and communication remains largely untapped.



Motivated by the aforementioned research gaps and opportunities, this paper proposes a novel \gls{ipsac} system framework to enable polarimetric sensing and improve the sensing-communication trade-off through the use of polarized signals.
\bb{We leverage single-port polarization-reconfigurable antenna arrays to benefit both communications and sensing from polarization diversity.
Compared to traditional dual-polarized \gls{mimo} that requires a dedicated RF chain for each polarization branch, this approach achieves polarization reconfigurability by combining two orthogonal linear polarization outputs through a single RF chain.
Consequently, this design significantly reduces hardware cost, complexity, and energy consumption by achieving polarimetric reconfigurability with only half the number of RF chains required by dual-polarized MIMO, as summarized in 
Table~\ref{tab:antenna_comparison}.
Our work develops a comprehensive signal processing framework, encompassing system modeling, optimization formulation, and algorithm design to fully exploit these advantages.
}

\bb{
\begin{table}[t!]
    \centering
    \caption{Comparison of polarized antenna architecture}
    \label{tab:antenna_comparison}
    \renewcommand{\arraystretch}{1.2} 
    \setlength{\tabcolsep}{4pt} 
    \begin{tabularx}{\columnwidth}{l >{\centering\arraybackslash}X >{\centering\arraybackslash}X >{\centering\arraybackslash}X}
        \toprule
        \textbf{Metric} & \textbf{Polarization-reconfigurable array} & \textbf{Single-polarized MIMO} & \textbf{Dual-polarized MIMO} \\
        \midrule
        \textbf{Number of RF chains}
        & $N_t$
        & $N_t$
        & $2N_t$
        \\
        \textbf{Power consumption}
        & Low
        & Lowest
        & High
        \\
        \textbf{Polarization DoF}
        & Medium
        & None (Static)
        & High
        \\
        \textbf{Implementation cost}
        & Low
        & Lowest
        & High \\
        \textbf{Polarimetric sensing}
        & Feasible
        & Infeasible
        & Feasible
        \\
        \bottomrule
    \end{tabularx}
\end{table}
}

\begin{figure*}[t]
\center{\includegraphics[width=.88\linewidth]{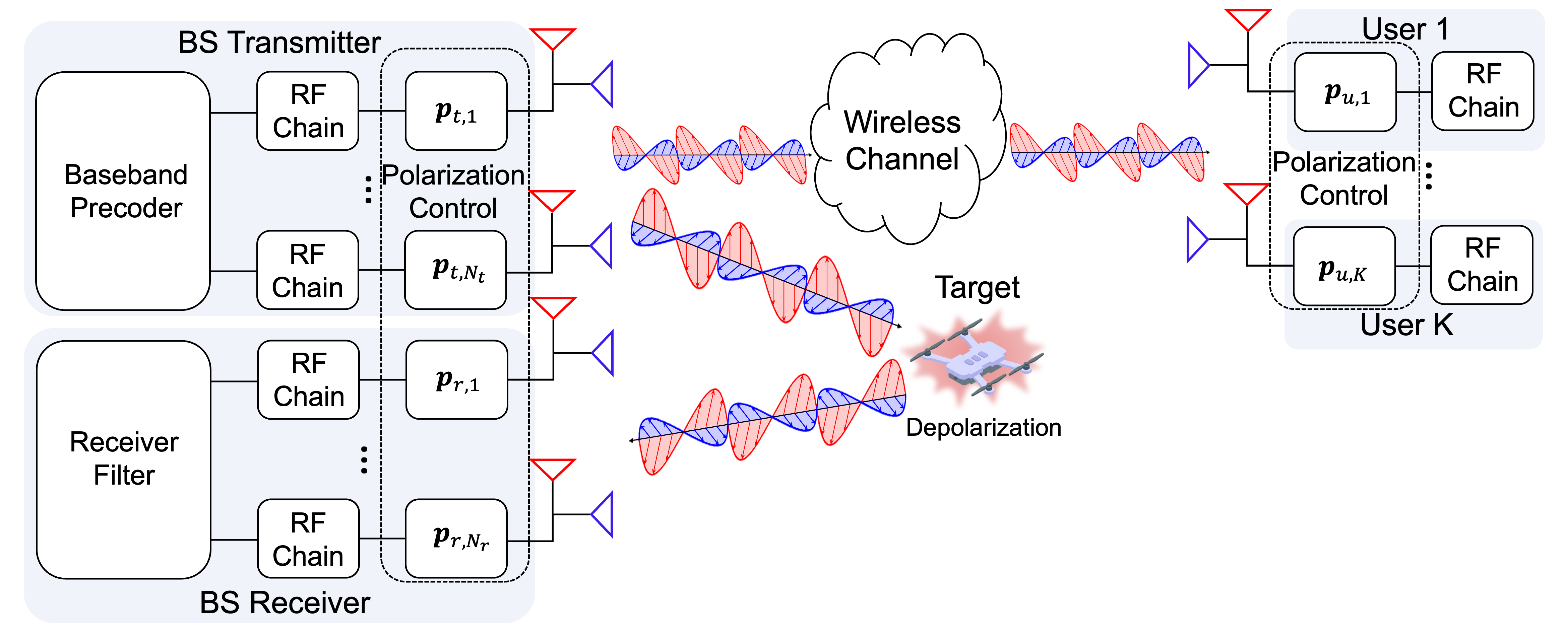}}
\caption{\small \gls{isac} system with polarization-reconfigurable arrays. Each antenna can adjust the linear polarizations of the transmitted and received signals.
\bb{The transmitted wave experiences depolarization at the target, and the backscattered signal is received at the BS receiver.}
}
\label{fig:system}
\vspace{-3mm}
\end{figure*}

\bb{
We address two primary tasks in \gls{ipsac} systems: depolarization parameter estimation and target detection.
Depolarization parameter estimation aims to infer the target's depolarization response, which provides information about the target's shape, orientation, and material composition.
Previous radar studies have addressed optimizing antenna polarization to improve the performance of depolarization estimation \cite{xiaoJointTransmitterReceiver2009,wangTargetPolarizationScattering2023}.
Building on this, we formulate a waveform and polarization optimization problem to minimize the depolarization parameter \gls{mse} for precise polarimetric sensing under per-user \gls{sinr} constraints. 
The polarization characteristics estimated at this stage are subsequently utilized in the target detection phase.
As prior polarimetric sensing studies \cite{novakStudiesTargetDetection1989,gogineniPolarimetricMIMORadar2010,demaioPolarizationDiversityDetection2004b,demaioPolarimetricAdaptiveDetection2002a} have demonstrated, exploiting such polarization statistics can significantly improve target detection and clutter suppression. 
Inspired by this, we incorporate polarization statistics into our waveform and polarization design to maximize the target \gls{sinr}.
Thus, depolarization parameter estimation and target detection are tightly coupled in our proposed IPSAC framework.
}
The main contributions are summarized as follows:
\begin{itemize}
    \item We establish a comprehensive signal model tailored for \gls{ipsac} systems using single-port polarization-reconfigurable antenna arrays. 
    This model allows the consideration of scenarios with multi-user communications and realistic sensing environments with interference from clutter and provides a foundation for the joint optimization of waveform and polarization configurations.
    \item 
    We formulate waveform and polarization optimization problems to minimize the \gls{mse} of the depolarization matrix estimate for precise and reliable polarimetric sensing.
    \bb{
    The resulting optimization problem has a nonconvex objective and unique per-antenna unit norm constraints, which we address by adopting \gls{sdr} for waveform optimization and the \gls{mm} approach for optimizing the linear polarizations.
    }
    \item 
    We extend our methodology to a practical target detection scenario, formulating and addressing a target \gls{sinr} maximization problem. 
    \bb{We exploit knowledge of the target and clutter polarization statistics to enhance target discrimination in the spatial and polarization domains. The resulting complex problem is efficiently solved by adapting our previously developed optimization framework.}

    \item 
    \bb{
    We perform extensive simulations comparing our proposed polarization-reconfiguration approach with traditional polarization-fixed and dual-polarized \gls{mimo} benchmarks.
    The results show that the proposed polarization-reconfiguration scheme consistently outperforms the static polarization baseline, while approaching the performance of the dual-polarized \gls{mimo} benchmark with only half the RF chains.
    These results confirm that the proposed \gls{ipsac} framework can effectively exploit polarization-domain degrees of freedom without increasing hardware costs, complexity, and power consumption.}    
\end{itemize}

\bb{
The rest of the paper is organized as follows. 
Section II proposes the signal and system models along with the sensing and communication performance metrics.
Sections III and IV respectively formulate the depolarization \gls{mse} minimization problem and the target \gls{sinr} maximization problem.
Section V evaluates the performance of the proposed waveform and polarization design algorithms, and finally
Section VI concludes the paper.
}

\textbf{Notation}:
Vectors and matrices are denoted by boldface lowercase and uppercase letters, respectively.
The superscripts $(\cdot)^T$, $(\cdot)^*$, $(\cdot)^H$, and $(\cdot)^{-1}$ indicate transpose, conjugate, conjugate transpose, and inverse.
Absolute value, two-norm, and Frobenius norm are denoted by $|\cdot|$, $\Vert\cdot\Vert$, and $\Vert\cdot\Vert_F$, respectively.
The operator $\operatorname{blkdiag}(\cdot)$ forms a block-diagonal matrix from its argument, while $\operatorname{vec}(\cdot)$ vectorizes a matrix.
Trace and rank are expressed by $\operatorname{Tr}(\cdot)$ and $\operatorname{Rank}(\cdot)$.
The real part of a complex number is extracted by $\operatorname{Re}(\cdot)$, and expectation is denoted by $\mathbb{E}[\cdot]$.
The $N\times N$ identity matrix is represented by $\textbf{I}_N$.
Kronecker and Hadamard products are indicated by $\otimes$ and $\odot$.
The relation $\textbf{A}\succeq\textbf{B}$ means $\textbf{A}-\textbf{B}$ is positive semidefinite, and the $(i,j)$-th entry of $\textbf{A}$ is denoted by $A_{i,j}$.




\section{System Model}

Consider an \gls{ipsac} system as illustrated in Fig. \ref{fig:system}, where a dual-functional \gls{bs} serves $K$ communication receivers while sensing a potential radar target. 
The primary goal of the system is to jointly optimize the waveform and linear polarization to efficiently perform polarimetric radar sensing, including target detection and depolarization parameter estimation, while communicating with multiple users. 
The \gls{bs} is equipped with colocated
transmit and receive arrays, consisting of $N_t$ transmit and $N_r$ receive antennas, respectively.
\bb{
Each BS transmit antenna has a single RF chain input connected to two orthogonally polarized antenna elements in the horizontal (H) and vertical (V) dimensions. 
The linear polarization applied at each antenna can be reconfigured by adjusting the relative amplitude of the signal flowing into each polarization branch.
In practice, this can be implemented using a simple analog combiner (e.g., a varactor-controlled combiner \cite{gaoPolarizationagileAntennas2006,caiContinuouslyTunablePolarization2017} or a variable attenuator \cite{zhou2024polarforming}) that controls the ratio of the amplitudes of the signals flowing into the H and V components.
Unlike conventional dual-polarized \gls{mimo}, this architecture achieves polarization reconfigurability without requiring a dedicated RF chain for each polarization branch.
}



The transmit array at the \gls{bs} radiates waveforms that simultaneously serve radar sensing and downlink communication users, while the colocated receive array captures radar echoes reflected from targets and clutter. 
Each communication receiver is equipped with a dual-polarized antenna pair to exploit polarization diversity at the user side.
The dual-function downlink transmit signal is given by 
\begin{equation}\label{eq:tx_signal}
    \textbf{X}=\textbf{F}_c\textbf{S}_c + \textbf{F}_s\textbf{S}_s,
\end{equation}
where $\textbf{F}_c\in\mathbb{C}^{N_t \times K}$ is the communication precoding matrix, $\textbf{F}_s\in\mathbb{C}^{N_t \times (N_t-K)}$ is the sensing precoding matrix, $\textbf{S}_c\in\mathbb{C}^{K \times L}$ is the communication signal, and $\textbf{S}_s\in\mathbb{C}^{(N_t-K) \times L}$ is the sensing signal.
The communication signal matrix $\textbf{S}_c$ has independent and identically distributed (i.i.d.) entries with zero mean and unit variance.
The sensing signal consists of $N_t-K$ orthogonal codewords of length $L$ with $L\geq N_t-K$, i.e., $\textbf{S}_s\textbf{S}_s^H=\textbf{I}_{N_t-K}$.
The transmit signal can be compactly written as 
$\textbf{X}=\textbf{F}\textbf{S}$ where $\textbf{F}=[\textbf{F}_c, \textbf{F}_s]$ and $\textbf{S}=[\textbf{S}_c^T,\textbf{S}_s^T]^T$.
The first $K$ columns $\textbf{f}_1,\textbf{f}_2,\dots,\textbf{f}_K$ in $\textbf{F}=[\textbf{f}_1,\textbf{f}_2,\dots,\textbf{f}_{N_t}]$ correspond to the beamformers for users $1,2,\dots,K$.
Let $\rho_t$ be the transmit power budget for the transmit signal, i.e., $\rho_t=\mathbb{E}[\Vert\textbf{X}\Vert_F^2]=\Vert\textbf{F} \Vert_F^2$.
Note that since $\textbf{S}_c\textbf{S}_c^H \rightarrow \textbf{I}_K$ as $L\rightarrow \infty$ and $\textbf{S}_s\textbf{S}_s^H=\textbf{I}_{N_t-K}$, then $\textbf{S}\textbf{S}^H \rightarrow \textbf{I}_{N_t}$. Hence $\textbf{R}_x \rightarrow \textbf{F}\textbf{F}^H$ as $L\rightarrow\infty$,
where $\textbf{R}_x=\textbf{X}\textbf{X}^H$ is the sample transmit covariance.
Based on the above, the asymptotic approximation $\textbf{R}_x\approx \textbf{F}\textbf{F}^H$
holds for sufficiently large $L$ \cite{liu2021cramer}.


The sensing and communication input-output relationships are respectively given by
\begin{equation}\label{eq:comm_sensing_sig}
    \begin{aligned}
        \textbf{Y}_s &=\textbf{T}\textbf{X}+\textbf{C}\textbf{X}+\textbf{W}_s\;,\\
        \textbf{Y}_c&=\textbf{H}\textbf{X}+\textbf{W}_c \; ,
    \end{aligned}
\end{equation}
where $\textbf{T}\in \mathbb{C}^{N_r \times N_t}$ is the target response matrix, $\textbf{C}\in \mathbb{C}^{N_r \times N_t}$ is the response due to clutter, $\textbf{H}\in \mathbb{C}^{N_u \times N_t}=[\textbf{h}_1,\textbf{h}_2,\dots,\textbf{h}_{N_u}]^H$ denotes the communication channel, $\textbf{h}_k$ is the channel for user $k$, $\textbf{W}_s\in \mathbb{C}^{N_r \times L}$ is white Gaussian noise at the 
\gls{bs} radar receiver with i.i.d. entries distributed as $\mathcal{CN}(0,\sigma_s^2)$, and $\textbf{W}_c\in \mathbb{C}^{N_u \times L}$ is white Gaussian noise at the users with i.i.d. entries distributed as $\mathcal{CN}(0,\sigma_c^2)$.
We assume that the target and clutter lie in the same ring of range cells which is known to the \gls{bs}.
Also, we consider a scenario where the target and clutter are stationary.
Hence we omit explicit representation of the delay and Doppler.
For communications, we assume a narrowband block-fading channel.
In what follows, we formalize models for the array response and the sensing and communication channels.

\subsection{Polarization-Reconfigurable Array Model}
To model the polarized \gls{mimo} channel, we adopt the array model in \cite{bhagavatulaNewDoubleDirectionalChannel2010,castellanosLinearPolarizationOptimization2024}.
An electromagnetic wave consists of an electric field and a magnetic field that propagate in the same direction but oscillate on perpendicular planes. 
The polarization of an electromagnetic wave is defined as the orientation of the electric field vector.
In particular, linear polarization refers to the case where the electric field vector oscillates in a single plane that maintains a constant orientation perpendicular to the direction of the propagating wave. 
Each transmit antenna launches a wave with a specific linear polarization, called the {\em co-polarization}, while the {\em cross-polarization} is orthogonal to the co-polarization and the direction of propagation.
When a radiated wave impinges on a scatterer, part of the wave energy may transfer to the cross-polarization component, changing its polarization state.
This phenomenon, known as \textit{depolarization}, is fundamental to modeling polarization behavior in wireless channels.

At the receiver, the antenna linearly combines the horizontal and vertical polarization components using a combining vector $\textbf{p}=[p_H,p_V]^T\in \mathbb{R}^{2}$, where $p_H,p_V$ denote the horizontal and vertical polarization gains, respectively, and $\Vert \textbf{p}\Vert ^2=p_H^2+p_V^2=1$. 
Ideally, the receive antennas are sensitive only to co-polarized signals, rejecting cross-polarized signals. However, in practice antennas may not perfectly separate the co- and cross-polar signals due to hardware imperfections.
To model such cross-polar leakage, we introduce a $2\times 2$ correlation matrix, called the antenna \gls{xpd} matrix, which is defined by \cite{oestgesDualpolarizedWirelessCommunications2008}
\begin{equation}
    \textbf{V} =\frac{1}{\sqrt{1+\chi_{ant}}}\begin{bmatrix}
        1 & \sqrt{\chi_{ant}} \\
        \sqrt{\chi_{ant}} &  1
    \end{bmatrix},
\end{equation}
where \bb{$\chi_{ant}$ is the antenna \gls{xpd} parameter that models the leakage between H and V elements}.
When $\chi_{ant}=0$, $\textbf{V}$ is a $2 \times 2$ identity matrix, representing perfect isolation between co- and cross-polarizations.
In contrast, when $\chi_{ant}=1$, $\textbf{V}$ becomes rank-one, corresponding to unpolarized antennas that cannot distinguish between co- and cross-polar signals.
In practice, polarized antennas are designed with $\chi_{ant} \ll 1$ to effectively discriminate polarized signals because the antenna \gls{xpd} $\chi_{ant}$ directly impacts the channel conditions and the sensing quality, e.g., the rank of the \gls{mimo} channel and the depolarization parameter \gls{mse}.
We assume the antenna \gls{xpd} matrix $\textbf{V}$ to be known.
To incorporate cross-polar leakage, the horizontal and vertical polarization gains can be modeled as
\begin{equation}\label{eq:pol_gain2}
\begin{aligned}
        [\hat{p}_H\ \hat{p}_V]
    &=\underbrace{[p_H\ p_V]}_{\textbf{p}^T}
    \textbf{V}
    =\textbf{p}^T\textbf{V}, \vspace{-3mm}
\end{aligned}    
\end{equation}
where $\hat{p}_H,\hat{p}_V$ are the effective horizontal and vertical polarization gains.

Next we formulate the array response based on the above polarization gain expressions. We will derive the array response model in a general way first, and then specialize the definition to the individual transmit and receive array models. Consider an array of $N$ antennas.
The \textit{net} array response matrix is defined as \cite{bhagavatulaNewDoubleDirectionalChannel2010,castellanosLinearPolarizationOptimization2024}
\begin{equation}
\begin{aligned}
    \textbf{A}_{net}(\theta)&=\left(\textbf{a}(\theta)\otimes [1\ 1]\right)\odot
    \begin{bmatrix}
        \hat{p}_{H,1} & \hat{p}_{H,2} & \ldots &\hat{p}_{H,N}\\
        \hat{p}_{V,1}  & \hat{p}_{V,2} & \ldots &\hat{p}_{V,N}
    \end{bmatrix}^T \\
    &= 
    \left(\textbf{a}(\theta)\otimes [1\ 1]\right)\odot
    \left([\textbf{p}_1,\textbf{p}_2,\ldots,\textbf{p}_N]^T\textbf{V}\right),
\end{aligned}
\end{equation}
where $\textbf{a}(\cdot)\in \mathbb{C}^N$ is the steering vector of the array, $\theta$ denotes either the \gls{aoa} or \gls{aod} depending on whether it refers to the receive or transmit side, and $\hat{p}_{H,n},\hat{p}_{V,n}$ represent the horizontal and vertical polarization gains for antenna $n$, respectively.
Note the parameter $\theta$ represents the azimuth and elevation angles for a two-dimensional array (e.g., a \gls{upa}), whereas it represents the azimuth angle only for a one-dimensional linear array (e.g., a \gls{ula}).
The second equality follows from $[\hat{p}_{H,n}, \hat{p}_{V,n}]=\textbf{p}^T_n\textbf{V}$, where $\textbf{p}_n$ denotes the polarization vector for antenna $n$.
Unlike conventional vector-form steering vectors, the \textit{net} array response matrix consists of two columns, with each column obtained by performing element-wise multiplication of the steering vector and corresponding horizontal or vertical polarization gains. 
This specific structure matches the polarization response of a wireless channel, expressed with respect to horizontal and vertical polarizations.

Polarization-reconfigurable arrays can tune the polarization vectors $\textbf{p}_1,\textbf{p}_2,\dots,\textbf{p}_N$ to mitigate polarization mismatch.
After some manipulations, the net array response matrix can be compactly rewritten as
\begin{equation}\label{eq:arr_resp_mat}
    \textbf{A}_{net}(\theta)
     =\textbf{P}^T\underbrace{\left(\textbf{a}(\theta)\otimes \textbf{V}\right)}_{\textbf{A}(\theta)}=\textbf{P}^T\textbf{A}(\theta) \; ,
\end{equation}
where $\textbf{P}=\text{blkdiag}(\textbf{p}_1,\textbf{p}_2,\dots,\textbf{p}_N)$.
From~\eqref{eq:arr_resp_mat}, the net transmit and receive array responses can be obtained as
\begin{equation}
    \textbf{A}_{net,t} =\textbf{P}^T_t\textbf{A}_t(\theta), \quad  
        \textbf{A}_{net,r} =\textbf{P}^T_r\textbf{A}_r(\theta),
\end{equation}
where the subscripts $t$ and $r$ on $\textbf{A}$ and $\textbf{P}$, respectively, denote the corresponding transmit and receive array parameters.
\bb{
Note that polarization-reconfigurable arrays coherently combine orthogonal polarization components before the RF chain, inherently boosting effective SNR. Moreover, the tunable polarization parameters $\textbf{P}_t$ and $\textbf{P}_r$ enable the formation of more favorable effective channels, further enhancing signal quality and suppressing interference.
}

\subsection{Sensing Signal Model}

Next we establish the target and clutter responses based on the array model in \eqref{eq:arr_resp_mat}.
Consider a single target at angle $\theta_0$.
From \eqref{eq:arr_resp_mat}, the target response is given by
\begin{equation}\label{eq:target_response}
\begin{aligned}
    \textbf{T}&=\beta_0 \textbf{A}_{net,r}(\theta_0)\bm{\Phi}_0\textbf{A}^T_{net,t}(\theta_0)\\
    &=\beta_0 \textbf{P}^T_r{\textbf{A}_r(\theta_0)\bm{\Phi}_0\textbf{A}^T_t(\theta_0)}\textbf{P}_t,
\end{aligned}
\end{equation}
where $\beta_0\in \mathbb{R}$ is the propagation loss through the medium (i.e., air), and $\bm{\Phi}_0$ is the $2\times 2$ target depolarization matrix.
The propagation loss $\beta_0$ can be modelled as $\beta_0=\eta/r^2$ where $\eta$ is a constant that captures system and environmental factors (e.g., carrier frequency, permittivity, and permeability) and $r$ is the radar-to-target distance.
Block-diagonal matrices $\textbf{P}_r=\text{blkdiag}\left(\textbf{p}_{r,1},\textbf{p}_{r,2},\dots,\textbf{p}_{r,N_r}\right)\in\mathbb{R}^{2N_r\times N_r}$ and $\textbf{P}_t=\text{blkdiag}\left(\textbf{p}_{t,1},\textbf{p}_{t,2},\dots,\textbf{p}_{t,N_t}\right)\in\mathbb{R}^{2N_t\times N_t}$ contain the receive and transmit polarization vectors, where $\textbf{p}_{r,n_r}$ denotes the receive polarization vector for antenna $n_r$ and $\textbf{p}_{t,n_t}$ denotes the transmit polarization vector for antenna $n_t$.
The target depolarization matrix is given by 
\begin{equation}
    \bm{\Phi}_0=\begin{bmatrix}
        \alpha^{HH}_0 & \alpha^{HV}_0 \\
        \alpha^{VH}_0 & \alpha^{VV}_0
    \end{bmatrix},
\end{equation}
where $\alpha^{ij}_0$ is the scattering coefficient from polarization $i$ to polarization $j$.
\bb{For example, $\alpha^{HV}_0$ represents the phase shift and attenuation from the $H$ to the $V$ polarization after the radiated signal interacts with the target.
The depolarization matrix, also known as the polarization scattering matrix, accounts for the scattering and depolarization effects at the target, i.e., the changes in the orientation and phase shifts of the polarized signals \cite{cloudeReviewTargetDecomposition1996}.}

Similarly, the clutter response matrix is given by
\begin{equation}\label{eq:clutter_response}
\begin{aligned}
    \textbf{C}
    &=\displaystyle\sum_{q=1}^{N_c}\beta_q \textbf{P}^T_r{\textbf{A}_r(\theta_q)\bm{\Phi}_q\textbf{A}^T_t(\theta_q)}\textbf{P}_t,
\end{aligned}
\end{equation}
where $N_c$ is the number of clutter patches, and $\{\beta_q, \theta_q, \bm{\Phi}_q\}$ are respectively the propagation loss, the angle, and the $2\times 2$ depolarization matrix for the $q$-th reflector.

From the above target and clutter response models, the sensing signal \eqref{eq:comm_sensing_sig} can be expanded as
\begin{equation}\label{eq:sensing_signal_mat}
\begin{aligned}
\textbf{Y}_s&=\underbrace{\beta_0 \textbf{P}^T_r{\textbf{A}_r(\theta_0)\bm{\Phi}_0\textbf{A}^T_t(\theta_0)}\textbf{P}_t\textbf{X}}_{\text{target response}} \\
&+
\underbrace{\displaystyle\sum_{q=1}^{N_c}\beta_q \textbf{P}^T_r{\textbf{A}_r(\theta_q)\bm{\Phi}_q\textbf{A}^T_t(\theta_q)}\textbf{P}_t\mathbf{X}}_{\text{clutter}}+\textbf{W}_s,
\end{aligned}
\end{equation}
which can further be converted to vector form
\begin{equation}\label{eq:sensing_signal}
\begin{aligned}
     \textbf{y}_s=\operatorname{vec}(\textbf{Y}_s) 
    &= \beta_0\textbf{M}\textbf{A}_0\pmb{\phi}_0+ 
    \displaystyle\sum_{q=1}^{N_c}\beta_q\textbf{M}\textbf{A}_q\pmb{\phi}_q+\textbf{w}_s,
\end{aligned}    
\end{equation}
where $\textbf{M}$ is the $LN_r \times 4N_tN_r$ measurement matrix with $\textbf{M}=\textbf{X}^T\textbf{P}^T_t\otimes \textbf{P}^T_r$, $\textbf{A}_0=\textbf{A}_t(\theta_0)\otimes \textbf{A}_r(\theta_0)$, $\textbf{A}_q=\textbf{A}_t(\theta_q)\otimes \textbf{A}_r(\theta_q)$ for all $q$, $\pmb{\phi}_0=\operatorname{vec}(\bm{\Phi}_0)$, $\pmb{\phi}_q=\operatorname{vec}(\bm{\Phi}_q)$ is the vectorized depolarization matrix for target $q$, and $\textbf{w}_{s}=\operatorname{vec}(\textbf{W}_{s})$.
Using properties of the Kronecker product, the measurement matrix can be represented as
\begin{equation}
    \textbf{M}=\bar{\textbf{X}}^H\bar{\textbf{P}}^T,
\end{equation}
where $\bar{\textbf{X}}= \textbf{X}^* \otimes \textbf{I}_{N_r}$ and $\bar{\textbf{P}}=\textbf{P}_t\otimes \textbf{P}_r$.


Let $\bm{\Sigma}_0=\mathbb{E}[\pmb{\phi}_0\pmb{\phi}_0^H]$ be the covariance of the depolarization parameter. 
From this, the covariance of the target response matrix can be expressed as $\bm{\Omega}_0 = \beta_0^{2}\textbf{A}_0\bm{\Sigma}_0{\textbf{A}}^H_0$.
Similarly, define $\bm{\Omega}_c=\mathbb{E}\left[\sum_{q=1}^{N_c}\sigma_q^2\textbf{A}_q\pmb{\phi}_q\pmb{\phi}_q^H\textbf{A}^H_q\right]$ as the covariance of the clutter response matrix where $\sigma_q^2=\mathbb{E}[|\beta_q|^2]$.
The covariance of the received signal is given by
$\mathbb{E}\left[\textbf{y}_s\textbf{y}_s^H\right]=\textbf{M}\bm{\Omega}\textbf{M}^H+\sigma_s^2\textbf{I}_{LN_r}$
where 
$\bm{\Omega}=\bm{\Omega}_0+\bm{\Omega}_c$.
We assume the \gls{bs} has knowledge of the target parameters\footnote{
In many applications, the radar already has information about the distance to the target and the characteristics of the medium enabling precise estimation of the target parameters \cite{xiaoJointTransmitterReceiver2009,gogineniPolarimetricMIMORadar2010}.
Moreover, environmental knowledge can be attained through a cognitive radar framework \cite{guerciCognitiveRadarKnowledgeaided2010}.
} including the propagation loss $\beta_0$, the covariance $\bm{\Sigma}_0$ of the target depolarization parameter and the covariance $\bm{\Omega}_c$ of the clutter.
\bb{
The target and clutter response matrices $\textbf{T}$ and $\textbf{C}$ can be shaped by tuning the per-antenna polarization vectors in $\textbf{P}_r$ and $\textbf{P}_t$ and leveraging knowledge of the target and clutter covariances $\bm{\Omega}_0$ and $\bm{\Omega}_c$.
}
Based on this, we define two sensing metrics based on the formulated sensing signal model in the following.

\subsubsection{Depolarization parameter \gls{mse}}
The depolarization parameter $\pmb{\phi}_0$ describes the depolarization effect (e.g., polarization orientation change) at the target, and thus 
estimating $\pmb{\phi}_0$ is a critical task in polarimetric sensing applications such as rainfall forecasting \cite{notarosPolarimetricWeatherRadar2022b,torvikClassificationBirdsUAVs2016,lee2017polarimetric}.
We consider \gls{lmmse} estimation of the target depolarization parameter, which minimizes the \gls{mse} using the linear combiner $\textbf{G}$:
\begin{equation}\label{eq:MSE}
    \text{MSE}_{\pmb{\phi}_0}=\min\mathbb{E}\left[\Vert \pmb{\phi}_0 - \textbf{G}^H\textbf{y}_s\Vert^2\right].
\end{equation}
The solution to~\eqref{eq:MSE} is given by \cite{kay1993fundamentals}
\begin{equation}\label{eq:lmmse_recv}
    \textbf{G}_{opt}=  \beta_0(\textbf{M}\bm{\Omega}\textbf{M}^H+\sigma_s^2\textbf{I}_{LN_r})^{-1}\textbf{M}\textbf{A}_0\bm{\Sigma}_0 ,
\end{equation}
with error covariance
\begin{equation}\label{eq:error_cov1}
    \begin{aligned}
        \textbf{Z}_e&={\bm{\Sigma}}_0-\beta_0^2{\bm{\Sigma}}_0{\textbf{A}}^H_0\textbf{M}^H\left(\textbf{M}\bm{\Omega}\textbf{M}^H+\sigma_s^2\textbf{I}_{LN_r} \right)^{-1}\textbf{M}\textbf{A}_0{\bm{\Sigma}}_0
    \end{aligned}
\end{equation}
and \gls{mse} given by $\text{MSE}_{\pmb{\phi}_0}(\textbf{M}) =\operatorname{Tr}(\textbf{Z}_e)$.

\subsubsection{Target \Gls{sinr}}
In addition to the target depolarization parameter \gls{mse}, we consider the target \gls{sinr}, which determines the target detection probability.
From the signal model \eqref{eq:sensing_signal}, the target \gls{sinr} is given by 
\begin{equation}\label{eq:metric:sinr}
\begin{aligned}
     \gamma_s(\textbf{M})
     &= \frac{\mathbb{E}\big[\operatorname{Tr}\big(\textbf{M}\bm{\Omega}_0\textbf{M}^H\big)\big]}{\mathbb{E}\big[\operatorname{Tr}\big(\textbf{M}\bm{\Omega}_c\textbf{M}^H\big)\big]+\sigma^2_s}\\
     &= \frac{\operatorname{Tr}\big(\bar{\textbf{P}}^T\bm{\Omega}_0\bar{\textbf{P}}\bar{\textbf{R}}_x\big)}{\operatorname{Tr}\big(\bar{\textbf{P}}^T\bm{\Omega}_c\bar{\textbf{P}}\bar{\textbf{R}}_x\big)+\sigma^2_s}\\
     &= \frac{\operatorname{Tr}\big(\bar{\textbf{F}}^H\bar{\textbf{P}}^T\bm{\Omega}_0\bar{\textbf{P}}\bar{\textbf{F}}\big)}{\operatorname{Tr}\big(\bar{\textbf{F}}^H\bar{\textbf{P}}^T\bm{\Omega}_c \bar{\textbf{P}}\bar{\textbf{F}}\big)+\sigma_s^2},
\end{aligned}
\end{equation}
where
\begin{equation}
    \begin{aligned}
        \bar{\textbf{R}}_x =\textbf{R}_x^* \otimes \textbf{I}_{N_r} \text{ and }
        \bar{\textbf{F}}=\textbf{F}^*\otimes \textbf{I}_{N_r}.
    \end{aligned}
\end{equation}

\subsection{Communication Signal Model}

From \eqref{eq:comm_sensing_sig}, the signal due to the $\ell$-th symbol at user $k$ is given by
\begin{equation}
        {y}_{c,k,\ell} = \textbf{h}_k^H\textbf{f}_k{s}_{k,\ell}+\displaystyle\sum_{\substack{k'\neq k}} \textbf{h}_k^H\textbf{f}_{k'}{s}_{k',\ell} + {w}_{c,k,\ell}, 
\end{equation}
where $\textbf{f}_k$ is the $k$th column of the precoder $\textbf{F}$, $s_{k,\ell}$ is the $\ell$-th data symbol for user $k$ and ${w}_{c,k,\ell}$ is noise distributed as ${w}_{c,k,\ell} \sim \mathcal{CN}({0},\sigma_c^2)$.
The communication channel for user $k$ is given by \cite{castellanosLinearPolarizationOptimization2024}
\begin{equation}
    \textbf{h}_k = \textbf{p}_{u,k}^T\underbrace{\displaystyle\sum_{m=1}^{M_k}\beta_{k,m} \textbf{V}_u\bm{\Phi}_{k,m}\textbf{Q}(\psi_k)\textbf{A}^T_t(\theta_{k,m})}_{\textbf{H}_{up,k}}\textbf{P}_t=\textbf{p}_{u,k}^T\textbf{H}_{up,k}\textbf{P}_t\;,
\end{equation}
where $\textbf{H}_{up,k}$ is the channel before applying the polarization vectors, $M_k$ is the number of multipaths, $\beta_{k,m}$ is the propagation loss, 
$\mathbf{p}_{u,k}$ denotes the receive polarization vector at the $k$-th user, 
$\textbf{V}_u$ is the user antenna \gls{xpd} matrix,
$\theta_{k,m}$ is the \gls{aod} of the $m$th path, $\bm{\Phi}_{k,m}$ is the depolarization matrix, and $\textbf{Q}(\psi_k)$ is the $2 \times 2 $ rotation matrix with rotation angle $\psi_k$.
The communication channel has the same structure as the target and clutter responses in \eqref{eq:target_response} and \eqref{eq:clutter_response}, except for an extra factor $\mathbf{Q}(\psi_k)$, which accounts for the polarization--orientation mismatch between the BS and user~$k$ due to the user's random orientation~$\psi_k$.
We assume the \gls{bs} has knowledge of the communication channel, which can be acquired through pilot transmissions with horizontal and vertical polarizations applied separately.
Based on the above model, the \gls{sinr} for user $k$ is given by
\begin{equation}
    \gamma_k = \frac{|\textbf{h}_k^H\textbf{f}_k|^2}{\sum_{\substack{k'\neq k}}|\textbf{h}_k^H\textbf{f}_{k'}|^2+\sigma_c^2}.
\end{equation}


\section{Depolarization MSE Minimization}
In this section, we study the problem of minimizing the \gls{mse} of the depolarization parameter estimate.
Since a target's depolarization parameter conveys details about the target's shape, pose, and reflectivity, its accurate estimation is critical to various polarimetric sensing applications like rainfall forecasting, target classification, material identification, and radar imaging \cite{notarosPolarimetricWeatherRadar2022b,torvikClassificationBirdsUAVs2016,lee2017polarimetric}.
We therefore propose a joint waveform and polarization design strategy with the aim of minimizing the estimation error of the target depolarization parameter.


\subsection{Problem Formulation}

Based on the previous discussions, we aim to accurately estimate the target depolarization parameter $\pmb{\phi}_0$ from the received signals. To achieve this, we jointly optimize the transmit waveform and the transmit, receive, and user polarizations to minimize the \gls{mse} of the \gls{lmmse} estimator for $\pmb{\phi}_0$.
Define the concatenated transmit and receive polarization vectors as $\textbf{p}_t=[\textbf{p}_{t,1}^T,\textbf{p}_{t,2}^T,\dots,\textbf{p}_{t,N_t}^T]^T$ and $\textbf{p}_r=[\textbf{p}_{r,1}^T,\textbf{p}_{r,2}^T,\dots,\textbf{p}_{r,N_r}^T]^T$, respectively.
The optimization problem is formulated as
\begin{mini!}|s| 
{\textbf{p}_t,\textbf{p}_r,\textbf{F}}{\text{MSE}_{\pmb{\phi}_0}(\textbf{M})}
{}{}\label{prob:isac}
\addConstraint{\gamma_k\geq \gamma_{th},~\forall k=1,2,\ldots,K}{} \label{ineq:sinr}
\addConstraint{\operatorname{Tr}(\textbf{F}\textbf{F}^H)=\rho_t}{}\label{ineq:total_power}
\addConstraint{\Vert\textbf{p}_{r,n_r}\Vert=1,\ \forall n_r =1,2,\dots,N_r}{}\label{ineq:unit_sphere2}
\addConstraint{\Vert\textbf{p}_{t,n_t}\Vert=1, \ \forall n_t =1,2,\dots,N_t }{}, \label{ineq:unit_sphere3}
\end{mini!}
where 
\eqref{ineq:sinr} is the per-user \gls{sinr} constraint with minimum SINR threshold $\gamma_{th}$, \eqref{ineq:total_power} is the transmit power constraint, and \eqref{ineq:unit_sphere2} and \eqref{ineq:unit_sphere3} are the unit-norm constraints on the polarization vectors. 

After removing irrelevant terms, problem \eqref{prob:isac} can be simplified as 
\begin{equation}\begin{aligned}
& \underset{\textbf{p}_r,\textbf{p}_t,\textbf{F}}{\min}-\operatorname{Tr}\big({\bm{\Sigma}}_0{\textbf{A}}^H_0\textbf{M}^H\left(\textbf{M}\bm{\Omega}\textbf{M}^H\!+\!\sigma_s^2\textbf{I} \right)^{-1}\textbf{M}\textbf{A}_0{\bm{\Sigma}}_0\big)\\
&\qquad \text{s.t.}\quad\eqref{ineq:sinr}-\eqref{ineq:unit_sphere3}.
\label{prob:mmse}
\end{aligned}\end{equation}
Directly solving \eqref{prob:mmse} is difficult due to the nonconvex objective and the unit-norm constraints on the polarization vectors. Hence, we will adopt an alternating optimization where the waveform and the polarizations are updated alternatingly.

\subsection{Waveform Optimization Algorithm}
We first optimize the precoder $\textbf{F}$ with the polarization vectors fixed.
Define $\bar{\bm{\Omega}}=\bar{\textbf{P}}^T\bm{\Omega}\bar{\textbf{P}}$ and assume $\bar{\bm{\Omega}}$ is invertible\footnote{Matrix $\bar{\bm{\Omega}}$ is positive semidefinite but may not be invertible depending on the rank of the target and clutter covariances. 
In case $\bar{\bm{\Omega}}$ is rank deficient, we can modify $\bar{\bm{\Omega}}$ as $\bar{{\bm{\Omega}}}_{new}=\bar{\bm{\Omega}}_{old}+\kappa\textbf{I}$, where $\kappa > 0$ is a small value, without a significant impact on performance.}.
Let $\bar{\bm{\Omega}}=\textbf{L}\textbf{L}^H$ be the Cholesky decomposition of $\bar{\bm{\Omega}}$.  
Then, it follows that
\begin{equation}\label{eq:inv_lemma}
    \begin{aligned}
    &\bar{\textbf{X}}\big(\bar{\textbf{X}}^H\bar{\bm{\Omega}}\bar{\textbf{X}}+\sigma_s^2\textbf{I}_{N_tN_r}\big)^{-1}\bar{\textbf{X}}^H\\
        &= {\textbf{L}^{-H}}\big[\textbf{I}_{N_tN_r}-\sigma_s^2\big(\textbf{L}^H\bar{\textbf{X}}\bar{\textbf{X}}^H\textbf{L}\!+\!\sigma_s^2\textbf{I}_{N_tN_r}\big)^{-1}\big]\textbf{L}^{-1}\\
        &= {\bar{\bm{\Omega}}^{-1}}-
        \sigma_s^2\bar{\bm{\Omega}}^{-1}\big(\bar{\textbf{X}}\bar{\textbf{X}}^H\!+\!\sigma_s^2\bar{\bm{\Omega}}^{-1}\big)^{-1}\bar{\bm{\Omega}}^{-1},
        \\ &= {\bar{\bm{\Omega}}^{-1}}-
        \sigma_s^2\bar{\bm{\Omega}}^{-1}\big(\textbf{R}_x^*\otimes \textbf{I}_{N_r}\!+\!\sigma_s^2\bar{\bm{\Omega}}^{-1}\big)^{-1}\bar{\bm{\Omega}}^{-1},
    \end{aligned}    
\end{equation}
where the first equality follows from the matrix inversion lemma \cite{golub2013matrix} and the last equality comes from $\bar{\textbf{X}}\bar{\textbf{X}}^H=\textbf{R}_x^*\otimes \textbf{I}_{N_r}$.
By substituting \eqref{eq:inv_lemma} into \eqref{prob:mmse}, the optimization problem can be reformulated as
\begin{mini}|s| 
{\textbf{R}_x\succeq 0}{\operatorname{Tr}\big(\bm{\Sigma}_0{\textbf{A}}^H_0\bar{\bm{\Omega}}^{-1}\left(\textbf{R}_x^*\otimes \textbf{I}_{N_r}\!+\sigma_s^2\bar{\bm{\Omega}}^{-1}\right)^{-1}\bar{\bm{\Omega}}^{-1}\textbf{A}_0\bm{\Sigma}_0\big)}
{}{}
\addConstraint{\operatorname{Tr}(\textbf{R}_x)=\rho_t}{}{}
\addConstraint{\gamma_k\geq \gamma_{th},~\forall k}{}.
\end{mini}
Notice that the optimization is now done over $\textbf{R}_x$ instead of precoder $\textbf{F}$ and remains nonconvex since $\textbf{R}_x$ is inside a matrix inverse.
To address this, we introduce an auxiliary variable $\textbf{J}$ such that 
\begin{equation}\label{eq:aux_U}
    \textbf{J}\succeq \bm{\Sigma}_0{\textbf{A}}^H_0\bar{\bm{\Omega}}^{-1}\left(\textbf{R}_x^*\otimes \textbf{I}_{N_r}\!+\sigma_s^2\bar{\bm{\Omega}}^{-1}\right)^{-1}\bar{\bm{\Omega}}^{-1}\textbf{A}_0\bm{\Sigma}_0.
\end{equation}
The Schur complement of \eqref{eq:aux_U} is given by
\begin{equation}\label{eq:Schur comp}
    \begin{bmatrix}
    \textbf{J} & \bm{\Sigma}_0\textbf{A}^H_0\bar{\bm{\Omega}}^{-1} \\
    \bar{\bm{\Omega}}^{-1}\textbf{A}_0\bm{\Sigma}_0 & \textbf{R}_x^*\otimes \textbf{I}_{N_r}\!+\sigma_s^2\bar{\bm{\Omega}}^{-1}
\end{bmatrix}\succeq 0.
\end{equation}

Next, we transform the \gls{sinr} constraints into linear matrix inequalities (LMIs).
Let $\textbf{R}_k=\textbf{f}_k\textbf{f}_k^H$ for $k=1,2,\dots,K$.
The per-user \gls{sinr} constraint can be expressed as
\begin{equation}
\begin{aligned}
    \frac{\textbf{h}_k^H\textbf{R}_k\textbf{h}_k}{\sum_{\substack{k'\neq k}}\textbf{h}_k^H\textbf{R}_{k'}\textbf{h}_k+\sigma_c^2}
    =
    \frac{\textbf{h}_k^H\textbf{R}_k\textbf{h}_k}{\textbf{h}_k^H(\textbf{R}_x-\textbf{R}_{k})\textbf{h}_k+\sigma_c^2} \geq \gamma_{th},
\end{aligned}
\end{equation}
where equality holds because $\textbf{R}_x-\textbf{R}_k=\sum_{k'\neq k } \textbf{R}_{k'}$.
because $\textbf{R}_x=\sum_{k=1}^K \textbf{R}_k$.
Note that here we use the asymptotic approximation $\textbf{R}_x\rightarrow\textbf{F}\textbf{F}^H$.
Multiplying by the denominator and moving the numerator to the right-hand side yields
\begin{equation}\label{eq:comm cons}
\begin{aligned}    
    \textbf{h}_k^H(\textbf{R}_x-(1+\gamma_{th}^{-1})\textbf{R}_{k})\textbf{h}_k+\sigma_c^2\leq  0.
\end{aligned}
\end{equation}
Based on this, the waveform optimization problem is reformulated as
\begin{mini}|s|
{\textbf{R}_x,\{\textbf{R}_k\}_{k=1}^K, \textbf{J}}{\operatorname{Tr}(\textbf{J})}{}{}
\addConstraint{
\begin{bmatrix}
    \textbf{J} & \bm{\Sigma}_0 \textbf{A}^H_0 \bar{\bm{\Omega}}^{-1} \\
    \bar{\bm{\Omega}}^{-1} \textbf{A}_0 \bm{\Sigma}_0 & \textbf{R}_x^*\otimes \textbf{I}_{N_r}\!
    + \sigma_s^2 \bar{\bm{\Omega}}^{-1}
\end{bmatrix}
\succeq 0
}
\addConstraint{\textbf{h}_k^H(\textbf{R}_x-(1+\gamma_{th}^{-1})\textbf{R}_{k})\textbf{h}_k+\sigma_c^2\leq  0\ \forall k}
\addConstraint{\textbf{R}_x \succeq \displaystyle\sum_{k=1}^K \textbf{R}_k,~{\textbf{R}}_k\succeq 0,~\ \forall k}
\addConstraint{\operatorname{Tr}(\textbf{R}_x) = \rho_t}
\addConstraint{\operatorname{Rank}(\textbf{R}_k)=1,~\ \forall k}.
\label{prob:sdr}
\end{mini}

We relax the rank-one constraint and transform the problem to  
a convex semidefinite program (SDP) that can be solved in polynomial time using standard convex solvers.
Once the optimal covariances are obtained, the precoding matrix must be recovered from the covariance.
The communication precoder can be obtained using the approach in \cite{liuJointTransmitBeamforming2020}, by defining
\begin{equation}\label{eq:recover_F}
\begin{aligned}
    \textbf{f}_k = (\textbf{h}_k^H\textbf{R}_k\textbf{h}_k)^{-\frac{1}{2}}\textbf{R}_k\textbf{h}_k,~\ \forall k.
\end{aligned}    
\end{equation}
The sensing precoder in \eqref{eq:tx_signal} can then be obtained through the Cholesky decomposition as $\textbf{F}_s\textbf{F}_s^H = \textbf{R}_x - \sum_{k=1}^K \textbf{f}_k\textbf{f}^H_k$.





   



\subsection{Polarization Optimization Algorithm}
\subsubsection{Update of $\textbf{p}_t$}
Next we optimize the transmit and receive polarization for the BS arrays for a given waveform $\textbf{X}$.
It is not straightforward to directly optimize the polarizations due to the complex structure of the objective along with the unit-norm constraint on each polarization vector. 
To address this, we first reformulate the objective and constraints with respect to the transmit polarization vector $\textbf{p}_t$.
Then, we approximate the objective with a linear majorizer and an inner convex approximation of the \gls{sinr} constraint.
Based on the strong duality of the reformulated problem, we find the optimal solution by solving the dual problem of the approximated problem via bisection.

We begin by majorizing the objective with a linear function based on the following lemma, whose proof can be found in Appendix \ref{sec:appendix_A}.

\begin{lemma}\label{theorem:majorization}
Let $\bar{\textbf{P}}_{i}$ be $\bar{\textbf{P}}$ at iteration $i$.
The objective can be majorized as
    \begin{equation}
    \begin{aligned}
       & -\operatorname{Tr}\big({\bm{\Sigma}}_0{\textbf{A}}^H_0\textbf{M}^H\left(\textbf{M}\bm{\Omega}\textbf{M}^H+\sigma_s^2\textbf{I}_{LN_r} \right)^{-1}\textbf{M}\textbf{A}_0{\bm{\Sigma}}_0\big) \\ 
       &\leq  \operatorname{Tr}\big( \bar{\textbf{P}}^T\bm{\Lambda}_i\big) + const,
    \end{aligned}    
\end{equation}
where
\begin{equation*}
    \begin{aligned}
        \bm{\Lambda}_i &= 2\operatorname{Re}\big(\left(\bm{\Omega}\!-\!\lambda_{\bm{\Omega}}\textbf{I}_{4N_tN_r}\right)\bar{\textbf{P}}_i\bar{\textbf{X}}\textbf{E}_i^{-1}\textbf{B}_i
        \!-\!\textbf{A}_0\bm{\Sigma}_0\big)\textbf{B}_{i}^H\textbf{E}_{i}^{-1}\bar{\textbf{X}}^H,\\        \textbf{E}_i&=\bar{\textbf{X}}^H\bar{\textbf{P}}_i^T\bm{\Omega}\bar{\textbf{P}}_i\bar{\textbf{X}}+\sigma_s^2\textbf{I}_{LN_r},  \textbf{B}_i=\bar{\textbf{X}}^H\bar{\textbf{P}}_i^T\textbf{A}_0{\bm{\Sigma}}_0,
    \end{aligned}
\end{equation*}
and $\lambda_{\bm{\Omega}}$ is the maximum eigenvalue of matrix $\bm{\Omega}$.
\end{lemma}

By leveraging the block-diagonal structure of $\bar{\textbf{P}}$, we can further rewrite the right-hand side of the inequality in Lemma~\ref{theorem:majorization} with respect to the transmit polarization vector $\textbf{p}_t$.
\begin{lemma}\label{lemma:mse_majorized}
The right-hand side of the inequality in Lemma \ref{theorem:majorization} can be rearranged as
\begin{equation}
    \operatorname{Tr}\big( \bar{\textbf{P}}^T\bm{\Lambda}_i\big) =\textbf{p}_t^T\textbf{d}^{(i)},
\end{equation}
where $i$ is the iteration index, 
\begin{equation}
    \begin{aligned}  
    \textbf{d}^{(i)}&=\bm{\Theta}_t^T\bar{\textbf{I}}_t^T\operatorname{vec}\big(\bar{\textbf{P}}_r^T\bm{\Lambda}_i\big),\\    
    \bm{\Theta}_t&=\text{blkdiag}(\textbf{e}_{N_t,1},\textbf{e}_{N_t,2},\dots,\textbf{e}_{N_t,N_t})\otimes \textbf{I}_2,\\
    \bar{\textbf{I}}_t&=\textbf{I}_{N_t}\otimes(\textbf{K}_{N_r,2N_t} \otimes \textbf{I}_{N_r})(\textbf{I}_{2N_t} \otimes \operatorname{vec}(\textbf{I}_{N_r})),   
    \end{aligned}
\end{equation}
and $\textbf{e}_{N_t,i}$ is the $i$th column of the identity matrix $\textbf{I}_{N_t}$.
\end{lemma}
See Appendix \ref{appendix:proof_obj_pt} for the detailed proof.

Next, we rewrite the per-user \gls{sinr} constraint with respect to the transmit polarization vector $\textbf{p}_t$ using the following lemma.
\begin{lemma}\label{lemma:sinr_constraint}
    The per-user \gls{sinr} constraint can be transformed to 
    \begin{equation}
        \textbf{p}_t^T\bar{\bm{\Psi}}_k \textbf{p}_t +\gamma_{th}\sigma_c^2 \leq  0,
    \end{equation}
    where
    \begin{equation}
    \begin{aligned}
    \bar{\bm{\Psi}}_k 
    &= 
    \bm{\Theta}_t^T\bm{\Psi}_k\bm{\Theta}_t 
    ,\
        \bm{\Psi}_k = \Big(\gamma_{th}\displaystyle\sum_{\substack{k'\neq k}}^K\textbf{f}^*_{k'}\textbf{f}^T_{k'}-\textbf{f}^*_{k}\textbf{f}^T_{k} \Big) \otimes \tilde{\textbf{h}}_k\tilde{\textbf{h}}_k^H.
    \end{aligned}        
    \end{equation}
    
\end{lemma}

See Appendix \ref{appendix:proof_sinr_const} for the detailed proof.

From Lemmas \ref{lemma:mse_majorized} and \ref{lemma:sinr_constraint}, the polarization optimization problem can be reformulated as
\begin{mini}|s| 
{\textbf{p}_t}
{\textbf{p}_t^T\textbf{d}^{(i)}}
{}{}
\addConstraint{\textbf{p}_t^T\bar{\bm{\Psi}}_k \textbf{p}_t +\gamma_{th}\sigma_c^2 \leq  0,~ \ \forall k}
\addConstraint{\Vert\textbf{p}_{t,n_t}\Vert=1,~\ \forall n_t}.
\label{prob:mse:QCLP}
\end{mini}
It should be noted that $\bar{\bm{\Psi}}_k$ is not guaranteed to be positive definite.
Thus, \eqref{prob:mse:QCLP} can be seen as a nonconvex quadratically constrained linear program due to the nonconvex quadratic constraint and the unit-norm constraint.

To solve~\eqref{prob:mse:QCLP}, we find an inner convex subset of the feasible set corresponding to the SINR constraint.
We again use a first-order Taylor expansion to upper-bound the quadratic term in Lemma \ref{lemma:sinr_constraint} at point $\textbf{p}_t^{(i)}$ as
\begin{equation}\label{eq:L3 cvx}
\begin{aligned}
    \textbf{p}_t^T\bar{\bm{\Psi}}_k\textbf{p}_t \leq & 
    \lambda_{\bar{\bm{\Psi}}_k} N_t+ 2\textbf{p}_t^T(\bar{\bm{\Psi}}_k-\lambda_{\bar{\bm{\Psi}}_k} \textbf{I})\textbf{p}_t^{(i)} \\
    &+(\textbf{p}_t^{(i)})^T(\lambda_{\bar{\bm{\Psi}}_k} \textbf{I}-\bar{\bm{\Psi}}_k)\textbf{p}_t^{(i)},
\end{aligned}    
\end{equation}
where $\lambda_{\bar{\bm{\Psi}}_k}$ is the maximum eigenvalue of ${\bar{\bm{\Psi}}_k}$.
Using this inequality, the inner approximation of the \gls{sinr} constraint can be expressed as
\begin{equation}\label{ineq:sinr_pt}
    \textbf{p}_t^T\textbf{u}_k+r_k \leq 0,
\end{equation}
where
\begin{equation}\label{eq:ukrk}
    \begin{aligned}
 \textbf{u}_k &= 2 (\bar{\bm{\Psi}}_k-\lambda_{\bar{\bm{\Psi}}_k} \textbf{I})\textbf{p}_t^{(i)} ,
  \\
        r_k &=\gamma_{th}\sigma_c^2+\lambda_{\bar{\bm{\Psi}}_k} N_t +(\textbf{p}_t^{(i)})^T(\lambda_{\bar{\bm{\Psi}}_k} \textbf{I}_{2N_t}-\bar{\bm{\Psi}}_k)\textbf{p}_t^{(i)}.
    \end{aligned}
\end{equation}

Based on \eqref{ineq:sinr_pt}, the transmit polarization optimization problem can be transformed into a linear program with unit-norm constraints:
\begin{mini}|s| 
{\textbf{p}_t}
{g(\textbf{p}_t)}
{}{}
\addConstraint{g_k(\textbf{p}_t)\leq 0,~\ \forall k }{}
\addConstraint{\Vert\textbf{p}_{t,n_t}\Vert=1,~ \ \forall n_t}{},
\label{prob:lp_norm}
\end{mini}
where $g(\textbf{p}_t)=\textbf{p}_t^T\textbf{d}$ and $g_k(\textbf{p}_t)=\textbf{p}_t^T\textbf{u}_k+r_k$.
Note that hereafter we temporarily drop the iteration index for ease of notation.
Given $\textbf{p}_t = [\textbf{p}_{t,1}^T,\textbf{p}_{t,2}^T,\dots,\textbf{p}_{t,N_t}^T]^T$, the objective and linear constraints in \eqref{prob:lp_norm} can be rewritten as
\begin{equation}
    \begin{aligned}
      g(\textbf{p}_t)&=\sum\nolimits_{n_t=1}^{N_t}  \textbf{p}_{t,n_t}^T\textbf{d}_{n_t},\\
      g_k(\textbf{p}_t)&=\sum\nolimits_{n_t=1}^{N_t}  \textbf{p}_{t,n_t}^T\textbf{u}_{k,n_t}+r_k,~\ \forall k,
    \end{aligned}
\end{equation}
where $\textbf{d}_{n_t}$ and $\textbf{u}_{k,n_t}$ are the $n_t$-th $2\times 1$ subvectors of $\textbf{d}$ and $\textbf{u}_k$, respectively.
From this, the dual problem of \eqref{prob:lp_norm} is given by
\begin{equation}\label{eq:dual}
\begin{aligned}
& \underset{\bm{\mu}}{\sup}\ \underset{\{\textbf{p}_{t,n_t}\}_{n_t=1}^{N_t}}{\min}
& & \displaystyle\sum_{n_t=1}^{N_t}\textbf{p}_{t,n_t}^T\Big(\textbf{d}_{n_t}+\displaystyle\sum_{k=1}^K\mu_{k}\textbf{u}_{k,n_t}\Big)+\displaystyle\sum_{k=1}^K\mu_{k}{r}_{k} \\ 
& \text{s.t.}
& &  \Vert\textbf{p}_{t,n_t}\Vert=1,~\ \forall n_t \\
& & & \mu_{k} \geq 0 , \ \forall k,
\end{aligned}
\end{equation}
where $\bm{\mu}=[\mu_1,\mu_2,\dots,\mu_K]^T$ and $\mu_k$ is the Lagrange multiplier for constraint $g_k(\textbf{p}_t)\leq 0$. 
The inner problem of \eqref{eq:dual} has a closed-form solution 
\begin{equation}\label{eq:pt_sol}
    \textbf{p}_{t,n_t}(\bm{\mu})=
    -\frac{\textbf{d}_{n_t}+\sum_{k=1}^K\mu_{k}\textbf{u}_{k,n_t}}{\big\Vert \textbf{d}_{n_t}+\sum_{k=1}^K\mu_{k}\textbf{u}_{k,n_t} \big\Vert}.
\end{equation}

Previous work \cite{heQCQPExtraConstant2022} showed that strong duality holds for this class of problems if the following conditions are satisfied:
\begin{align}
    &0\leq \mu_k \leq \infty,~\ {g}_{k}(\bm{\mu})\leq 0, ~\forall k, \label{ineq:dual_feasible}\\
    &\mu_k{g}_{k}(\bm{\mu})=0, \forall k  \label{eq:KKT},
\end{align}
where ${g}_{k}(\bm{\mu})$ denotes $g_k(\textbf{p}_t(\bm{\mu}))$.
Eq.~\eqref{ineq:dual_feasible} represents the dual and primal feasible conditions and \eqref{eq:KKT} is the complementary slackness condition.
Thus, we aim to find the optimal solution that satisfies the above conditions.
Substituting $\textbf{p}_t$ in \eqref{eq:dual} with the closed-form solution \eqref{eq:pt_sol} yields 
\begin{equation}
\begin{aligned}
& \underset{\bm{\mu}}{\sup}
& &  -\displaystyle\sum_{n_t=1}^{N_t}\Big\Vert \textbf{d}_{n_t}+\displaystyle\sum_{k=1}^K\mu_{k}\textbf{u}_{k,n_t} \Big\Vert-\displaystyle\sum_{k=1}^K\mu_{k}r_{k} \\ 
& \text{s.t.}
& &   \mu_{k} \geq 0 , \ \forall k,\ {g}_{k}(\bm{\mu})\leq 0, ~\forall k,\\
 & & &\mu_k{g}_{k}(\bm{\mu})=0,~\forall k.
\end{aligned}
\end{equation}
To find the Lagrange multipliers satisfying the conditions \eqref{ineq:dual_feasible} and \eqref{eq:KKT}, we use a coordinate ascent method where one Lagrange multiplier is optimized via a line search while the others remain static \cite{heQCQPExtraConstant2022}.
This algorithm is referred to as a $K$-bisection search and is described in Algorithm \ref{alg:K-Bisection}.
Once the Lagrange multipliers are found, the polarization vector $\textbf{p}_{t,n_t}$ can be recovered using \eqref{eq:pt_sol}.

\setlength{\textfloatsep}{0pt}
\begin{algorithm}\small
\DontPrintSemicolon
\SetNlSty{textbf}{\{}{\}}
\SetAlgoNlRelativeSize{0}
\SetNlSty{}{}{}

\caption{
$K$-Bisection Search for finding $\bm{\mu}$ \label{alg:K-Bisection}}

\textbf{Input:} Lagrange multiplier vector {$\bm{\mu}$, stopping thresholds $\epsilon_1$, $\epsilon_2$}, $\epsilon_3$

\textbf{Initialization:} $i=0$; $\bm{\mu}[0] = \bm{\mu}$, $\hat{g}_{k}[0]=\infty$;
$g_k(\mu')$ denotes $g_k(\bm{\mu})|_{\mu_k=\mu'}$\;

\Repeat{$|\bar{g}[i]-\bar{g}[i-1]|/|\bar{g}[i-1]|<\epsilon_1$}{
    \SetKwBlock{DoParallel}{do in parallel}{end}
    \For{$k=1:K$}{
        \lIf{$g_k(0)\leq 0$}{$\mu_k=0$}
        \ElseIf{$\lim_{\mu_k \rightarrow \infty}|g_k(\mu_k)| \leq \epsilon_2$}{Stop Algorithm \ref{alg:K-Bisection}}
        \Else{
            $\mu^l=0$, $\mu^u=1$;\;
            \lIf{$g_k(\mu^u)\leq 0$}{$\mu^u=1$}
            
            \Else{
                \lRepeat{$g_k(\mu^u)\leq0$}{$\mu^u=2\mu^u$}
                $\mu^l=\mu^u/2$\;
            }
            \Repeat{$|g_k(\mu_k)+\epsilon_3/2|<\epsilon_3/2$}{
                $\mu_k=(\mu^l+\mu^u)/2$;\;
                \lIf{$g_k(\mu_k)>0$}{$\mu^l=\mu_k$}
                \lElse{$\mu^u=\mu_k$}
            }
        }
    }
    Update $i \gets i+1$,\ set $\bm{\mu}[i]=[\mu_1,\dots,\mu_{K}]$\;
    Update $\bar{g}[i]={g}_{\ell}(\bm{\mu}\;[i])+\sum_{k=1}^{K}\mu_kg_k(\bm{\mu}[i])$\;
}

{\textbf{Output:} Recover a solution $\textbf{p}_t$ from $\bm{\mu}[i]$ and \eqref{eq:pt_sol}}

\end{algorithm}

\setlength{\textfloatsep}{0pt}
\begin{algorithm}[t]\small 
\DontPrintSemicolon
\SetNlSty{textbf}{\{}{\}}
\SetAlgoNlRelativeSize{0}
\SetNlSty{}{}{}

\caption{Polarization Optimization Algorithm\label{alg:tx_pol}}

{\textbf{Input:} Initial point $\textbf{p}_t^{(0)},\textbf{p}_r^{(0)},\textbf{p}_u^{(0)}$, stopping threshold $\epsilon_4$}, maximum iteration $i_{max}$

\textbf{Initialize:} Set $i=0$, $\textbf{p}_t^{(i)}=\textbf{p}_t^{(0)}$, $\tilde{g}[i]=\infty$\;

\Repeat{$|\tilde{g}[i]-\tilde{g}[i-1]|/|\tilde{g}[i-1]|\leq \epsilon_4$ \text{or} $i\geq i_{max}$}{
    $i \gets i+1$\;
    Update $\textbf{p}_t^{(i)}$ using Algorithm \ref{alg:K-Bisection}\;
    Update $\textbf{p}_r^{(i)}$ and $\textbf{p}_u^{(i)}$ using \eqref{eq:update_pr} and \eqref{eq:update_pu}, respectively \;
    $\tilde{g}[i] \gets \text{MSE}_{\pmb{\phi}_0}(\textbf{p}_t^{(i)},\textbf{p}_r^{(i)},\textbf{p}_u^{(i)},\textbf{F})$\;
}

{\textbf{Output:} Polarization vectors $\textbf{p}_t$, $\textbf{p}_r$, and $\textbf{p}_u$\; }

\end{algorithm}

\subsubsection{Update of $\textbf{p}_r$}
Next, we optimize the receive polarization vectors for the radar receiver.
For a fixed $\textbf{p}_t$, the receive polarization can be optimized by solving 
\begin{mini}|s| 
{\textbf{p}_{r}}{\textbf{p}^T_{r}\textbf{q}^{(i)}}
{}{}
\addConstraint{\Vert\textbf{p}_{r,n_r} \Vert=1,~ \forall n_r}{},
\label{eq:pr}
\end{mini}
where 
$\textbf{q}^{(i)}=\bm{\Theta}_r^T\bar{\textbf{I}}_r^T\operatorname{vec}(\bar{\textbf{P}}_t^T\bm{\Lambda}_i)$. A proof for~\eqref{eq:pr} can also be found in Appendix \ref{appendix:proof_obj_pt}.
Since subvectors $\textbf{p}_{r,1},\textbf{p}_{r,2},\dots,\textbf{p}_{r,N_r}$ are independent, we can break the above problem into separate subproblems as
\begin{mini}|s| 
{\small{\textbf{p}^T_{r,n_r}:\Vert\textbf{p}^T_{r,n_r}\Vert=1}}{\textbf{p}^T_{r,n_r}\textbf{q}^{(i)}_{r,n_r}}
{}{},
\end{mini}
where $\textbf{q}_{r,n_r}$ is the $n_r$-th subvector of $\textbf{q}_r$.
The closed-form solution to each subproblem can be readily obtained as
\begin{equation}\label{eq:update_pr}
    \textbf{p}_{r,n_r}^{(i+1)}=\textbf{q}^{(i)}_{r,n_r}/\Vert \textbf{q}^{(i)}_{r,n_r} \Vert.
\end{equation}




\subsubsection{Update of $\textbf{p}_{u,k}$}\label{sec:update pu}
Although the user polarization vectors $\textbf{p}_{u,k}$ do not directly affect the sensing MSE objective, optimizing them can indirectly enhance the overall system performance. Specifically, by maximizing the communication SINR at the user side, we create additional DoFs in the subsequent waveform and polarization optimization steps, thereby indirectly reducing the MSE of the target depolarization parameter estimate. Thus, the optimization problem for updating $\textbf{p}_{u,k}$ is formulated as  
\begin{maxi}|s| 
{\textbf{p}_{u,k}}{\frac{\textbf{p}_{u,k}^T\hat{\textbf{H}}_k^{(i)}\textbf{p}_{u,k}}{\textbf{p}_{u,k}^T\bar{\textbf{H}}_k^{(i)}\textbf{p}_{u,k}}}
{}{}
\addConstraint{\Vert\textbf{p}_{u,k}\Vert =1}{},
\label{prob:user_sinr}
\end{maxi}
where $\hat{\textbf{H}}_k^{(i)}=\textbf{H}_{up,k}\textbf{P}_t^{(i)}\textbf{f}_k\textbf{f}_k^H(\textbf{P}_t^{(i)})^T\textbf{H}_{up,k}^H$ and $\bar{\textbf{H}}_k^{(i)}=\textbf{H}_{up,k}\textbf{P}_t^{(i)}\sum_{\substack{k' \neq k}}\textbf{f}_{k'}\textbf{f}_{k'}^H(\textbf{P}_t^{(i)})^T\textbf{H}_{up,k}^H+\sigma_c^2\textbf{I}_2$.
This problem can be equivalently recast as
\begin{maxi}|s| 
{\textbf{p}_{u,k}}{\textbf{p}_{u,k}^T\hat{\textbf{H}}_k^{(i)}\textbf{p}_{u,k}}
{}{}
\addConstraint{\textbf{p}_{u,k}^T\bar{\textbf{H}}_k^{(i)}\textbf{p}_{u,k} =1}{}.
\label{prob:user_sinr2}
\end{maxi}
The stationary condition of \eqref{prob:user_sinr2} is given by $(\bar{\textbf{H}}_k^{(i)})^{-1}\hat{\textbf{H}}_k^{(i)}\textbf{p}_{u,k}=\lambda_k \textbf{p}_{u,k}$ where $\lambda_k$ is the associated Lagrange multiplier \cite{boydConvexOptimization2004}.
Thus, the optimal value of $\textbf{p}_{u,k}$ can be obtained as
\begin{equation}\label{eq:update_pu}
    \textbf{p}_{u,k}^{(i)}=\textbf{t}^{(i)},
\end{equation}
where $\textbf{t}^{(i)}$ is the eigenvector of $(\bar{\textbf{H}}_k^{(i)})^{-1}\hat{\textbf{H}}^{(i)}_k$ corresponding to the maximum eigenvalue.



{\subsection{Summary}}
In the previous subsections, we developed waveform and polarization optimization algorithms to minimize the depolarization \gls{mse}.
A \gls{mse}-optimal waveform can be formulated by solving \eqref{prob:sdr} without the rank constraint and recovering the precoders from the calculated covariance using \eqref{eq:recover_F}.
The polarization vectors $\textbf{p}_t$, $\textbf{p}_r$, and $\textbf{p}_u$ can be obtained through iterative updates using Algorithm \ref{alg:K-Bisection}, \eqref{eq:update_pr}, \eqref{eq:update_pu}, respectively, until convergence, as summarized in Algorithm \ref{alg:tx_pol}.

\section{Target SINR Maximization}
In this section, we consider the problem of maximizing the target \gls{sinr}, a key metric that directly determines radar detection performance in cluttered environments, subject to the same communication SINR constraints and transmit power budget as in the previous section. 
Based on the formulated target \gls{sinr} in \eqref{eq:metric:sinr}, the optimization problem can be formulated as
\begin{maxi}|s| 
{\textbf{p}_r,\textbf{p}_t,\textbf{F}}
{\gamma_s(\textbf{M}) }
{}{}
\addConstraint{\eqref{ineq:sinr}\sim\eqref{ineq:unit_sphere3}.}{}
\label{prob:sinr}
\end{maxi}
Since this problem shares a similar structure with the depolarization MSE minimization problem studied in the previous section, we adopt a similar optimization framework and employ the previously developed \gls{sdr}- and \gls{mm}-based algorithms. In the following subsections, we describe the iterative solutions in detail. 

\subsection{Waveform Optimization Algorithm}
Given fixed polarizations, the waveform optimization problem is given by 
\begin{maxi}|s| 
{\textbf{R}_x}{\frac{\operatorname{Tr}\big(\bar{\textbf{P}}^T\bm{\Omega}_0\bar{\textbf{P}}\bar{\textbf{R}}_x\big)}{\operatorname{Tr}\big(\bar{\textbf{P}}^T\bm{\Omega}_c\bar{\textbf{P}}\bar{\textbf{R}}_x\big)+\sigma^2_s}}
{}{}
\addConstraint{\eqref{ineq:sinr}\sim\eqref{ineq:unit_sphere3}.}{}
\end{maxi} 
This problem is a fractional semi-definite program with respect to the  variable $\textbf{R}_x$.
Solving this fractional SDP by iteratively applying convex SDP solvers with a bisection method incurs high computational complexity \cite{boydConvexOptimization2004}.
To overcome this issue, we use the Charnes-Cooper transformation to convert the linear fractional program into an equivalent convex \gls{sdp} \cite{cooper1962programming}.
Specifically, we introduce the auxiliary variable ${t}=
(\operatorname{Tr}(\bar{\textbf{P}}^T\bm{\Omega}_c\bar{\textbf{P}}\bar{\textbf{R}}_x)+\sigma^2_s)^{-1}$ and define new variables  $\tilde{\textbf{R}}_x=t\textbf{R}_x$ and $\tilde{\textbf{R}}_k=t\textbf{R}_k$.
Based on these new definitions, the waveform optimization problem can be reformulated as 
\begin{maxi}|s| 
{t,\tilde{\textbf{R}}_x}{{\operatorname{Tr}\big(\bar{\textbf{P}}^T\bm{\Omega}_0\bar{\textbf{P}}(\tilde{\textbf{R}}_x^*\otimes \textbf{I}_{N_r})\big)}}
{}{}
\addConstraint{\operatorname{Tr}\big(\bar{\textbf{P}}^T\bm{\Omega}_c\bar{\textbf{P}}(\tilde{\textbf{R}}_x^*\otimes \textbf{I}_{N_r})\big)+t\sigma_s^2=1}
\addConstraint{\tilde{\textbf{R}}_x \succeq \displaystyle\sum_{k=1}^K \tilde{\textbf{R}}_k, ~\tilde{\textbf{R}}_k\succeq 0,~\ \forall k} 
\addConstraint{\textbf{h}_k^H\left(\tilde{\textbf{R}}_x-(1+\gamma_{th}^{-1})\tilde{\textbf{R}}_{k}\right)\textbf{h}_k+t\sigma_c^2\leq 0,~\ \forall k} 
\addConstraint{\operatorname{Tr}(\tilde{\textbf{R}}_x)=t\rho_t,~t>0},
\label{prob:sinr_sdp}
\end{maxi}
which is a convex \gls{sdp} and thus can be solved using standard optimization tools.
Once $\tilde{\textbf{R}}_x$ and $t$ are obtained, the transmit covariance can be recovered as $\textbf{R}_x = \tilde{\textbf{R}}_x/t$. Subsequently, the precoding matrix $\textbf{F}$ can be retrieved from $\textbf{R}_x$ following the same procedure as in the MSE minimization waveform design.

\subsection{Polarization Optimization Algorithm}
The fractional dependence of the target \gls{sinr} and the unit-norm constraint on the polarization vectors make direct maximization of the target \gls{sinr} challenging. 
To address this difficulty, we first apply Dinkelbach's  transform to convert the fractional objective into a polynomial form, and then utilize the \gls{mm} technique to construct a convex surrogate objective function. 
Specifically, by introducing an auxiliary variable $\nu$, which can be seen as the target SINR in the $i$th iteration, the sensing \gls{sinr} maximization problem can be reformulated as
\begin{mini}|s| 
{\textbf{p}_t,\textbf{p}_r}{\operatorname{Tr}\big(\bar{\textbf{F}}^H\bar{\textbf{P}}^T\bm{\Xi}_i\bar{\textbf{P}}\bar{\textbf{F}}\big)}
{}{}
\addConstraint{\eqref{ineq:sinr},~\eqref{ineq:total_power},~\eqref{ineq:unit_sphere2},~\eqref{ineq:unit_sphere3},}{}
\label{prob:sinr:pt}
\end{mini}
where $\bm{\Xi}_i=\nu(\bm{\Omega}_c+(\sigma^2_s/N_r)\textbf{I}_{N_rN_t})-\bm{\Omega}_0$.
Next, we construct a convex upper bound of the objective function at the point $\bar{\textbf{P}}_i$ via the first-order Taylor expansion as
\begin{equation}    
\begin{aligned}    &\operatorname{Tr}\big(\bar{\textbf{F}}^H\bar{\textbf{P}}^T\bm{\Xi}_i\bar{\textbf{P}}\bar{\textbf{F}}\big) \leq  2\operatorname{Tr}\big(\bar{\textbf{P}}^T(\bm{\Xi}_i-\lambda_{\bm{\Xi}_i}\textbf{I})\bar{\textbf{P}}_i\bar{\textbf{R}}_x\big)\\ &+\operatorname{Tr}\big(\bar{\textbf{P}}_i^T(\lambda_{\bm{\Xi}_i}\textbf{I}-\bm{\Xi}_i)\bar{\textbf{P}}_i\bar{\textbf{R}}_x\big)+\lambda_{\bm{\Xi}_i}N_r\rho_t,
\end{aligned}
\end{equation}
where $\lambda_{\bm{\Xi}_i}$ is the maximum eigenvalue of $\bm{\Xi}_i$. 
Then, following the same procedure as in Lemma 3 and \eqref{eq:L3 cvx}-\eqref{eq:ukrk}, we can also transform the communication SINR constraint into a linear function as in \eqref{ineq:sinr_pt}. With these convex surrogate functions established, we next iteratively update the polarization vectors $\textbf{p}_t$, $\textbf{p}_r$, $\textbf{p}_u$, and the auxiliary variable $\nu$, as detailed below.

\subsubsection{Update of $\textbf{p}_t$}
Fixing the other variables and ignoring irrelevant terms, the update for $\textbf{p}_t$ can be formulated as 
\begin{mini}|s| 
{\textbf{p}_t}{\textbf{p}^T_t \textbf{g}^{(i)}}
{}{}
\addConstraint{\textbf{p}_t^T\textbf{u}_k+r_k \leq 0,~\forall k}{} 
\addConstraint{\Vert\textbf{p}_{t,n_t}\Vert=1,~\ \forall n_t,}
\label{prob:sinr_lp}
\end{mini}
where $\textbf{g}^{(i)}=\bm{\Theta}_t^T\bar{\textbf{I}}_t^T\operatorname{vec}\big((\bar{\textbf{P}}_r)^T(\bm{\Xi}_i-\lambda_{\bm{\Xi}_i} \textbf{I})\bar{\textbf{P}}_i\bar{\textbf{R}}_x\big)$.
The proof of this result can be found in Appendix \ref{appendix:proof_sinr_majorizer}.
Note that hereafter we temporarily drop the iteration index for ease of notation.
Recall that the above problem formulation is equivalent to linear programming with unit-norm constraints on the subvectors of $\textbf{p}_t$.
As in the previous section, we solve the dual problem of \eqref{prob:sinr_lp} to find the optimal solution, which is given by
\begin{equation}\label{eq:dual_sinr}
\begin{aligned}
& \underset{\bm{\mu}}{\sup}\ \underset{\{\textbf{p}_{t,n_t}\}_{n_t=1}^{N_t}}{\min}
& & \displaystyle\sum_{n_t=1}^{N_t}\textbf{p}_{t,n_t}^T\Big(\textbf{g}_{n_t}+\displaystyle\sum_{k=1}^K\mu_{k}\textbf{u}_{k,n_t}\Big)+\displaystyle\sum_{k=1}^K\mu_{k}{r}_{k} \\ 
& \text{s.t.}
& &  \Vert\textbf{p}_{t,n_t}\Vert=1,~\ \forall n_t, \\
& & & \mu_{k} \geq 0 ,~ \ \forall k,
\end{aligned}
\end{equation}
where $\bm{\mu}=[\mu_1,\mu_2,\dots,\mu_K]^T$, $\textbf{g}_{n_t}$ is the $n_t$-th $2\times 1$ subvector of $\textbf{g}$, and $\mu_k$ is the Lagrange multiplier for the $k$-th user's communication constraint.
The inner problem has a closed-form solution
\begin{equation}\label{eq:pt_sol_sinr}
    \textbf{p}_{t,n_t}(\bm{\mu})=
    -\frac{\textbf{g}_{n_t}+\sum_{k=1}^K\mu_{k}\textbf{u}_{k,n_t}}{\big\Vert \textbf{g}_{n_t}+\sum_{k=1}^K\mu_{k}\textbf{u}_{k,n_t} \big\Vert}.
\end{equation}
As in the previous section, the outer problem can be solved by finding the Lagrange multipliers using the $K$-bisection search in Algorithm \ref{alg:K-Bisection}, and the polarization vector can be recovered from \eqref{eq:pt_sol_sinr}.

\subsubsection{Update of $\textbf{p}_r$}
Given the other variables, the receive polarization can be optimized by solving 
\begin{mini}|s| 
{\textbf{p}_{r}}{\textbf{p}^T_{r}\textbf{v}^{(i)}}
{}{}
\addConstraint{\Vert\textbf{p}_{r,n_r} \Vert=1,~ \forall n_r}{},
\end{mini}
where $\textbf{v}^{(i)}=\bm{\Theta}_r^T\bar{\textbf{I}}_r^T\operatorname{vec}((\bar{\textbf{P}}_t)^T(\bm{\Xi}_i-\lambda_{\bm{\Xi}_i} \textbf{I})\bar{\textbf{P}}_i\bar{\textbf{R}}_x)$ (see Appendix \ref{appendix:proof_sinr_majorizer} for the proof).
The closed-form solution to each subproblem can be readily obtained as
\begin{equation}\label{eq:update_pr2}
    \textbf{p}_{r,n_r}^{(i+1)}=\textbf{v}^{(i)}_{r,n_r}/\Vert \textbf{v}^{(i)}_{r,n_r} \Vert ,
\end{equation}
where $\textbf{v}^{(i)}_{r,n_r}\in\mathbb{R}^{2\times 1}$ is the $i$th subvector of $\textbf{v}^{(i)}$.

\subsubsection{Update of $\textbf{p}_u$}
The update of the polarization vector $\mathbf{p}_u$ at the user side is identical to that in Sec. \ref{sec:update pu}.

\subsubsection{Update of auxiliary variable $\nu$}
Once the polarization vectors are found, the auxiliary variable $\nu$ can be updated as 
\begin{equation}
    \nu=\frac{\operatorname{Tr}\big(\bar{\textbf{F}}^H\bar{\textbf{P}}^T\textbf{A}_0\bm{\Sigma}_0\textbf{A}_0^H\bar{\textbf{P}}\bar{\textbf{F}}\big)}{\operatorname{Tr}\big(\bar{\textbf{F}}^H\bar{\textbf{P}}^T\bm{\Omega}_c\bar{\textbf{P}}\bar{\textbf{F}}\big)+\sigma_s^2}.
\end{equation}

{\subsection{Summary}}
In summary, the target-\gls{sinr}-optimal waveform can be obtained by solving \eqref{prob:sinr_sdp} and retrieving the precoder from the optimized covariance using
\eqref{eq:recover_F}.
Similar to the depolarization \gls{mse} minimization, the polarization vectors can be optimized by iteratively updating $\textbf{p}_t$, $\textbf{p}_r$, and $\textbf{p}_u$ using \eqref{eq:pt_sol_sinr} along with Algorithm \ref{alg:K-Bisection}, \eqref{eq:update_pr2}, and \eqref{eq:update_pu}, respectively.








\section{Numerical Results}
\subsection{Simulation Setup}


In this section, we evaluate the proposed approach through extensive simulations.
Unless otherwise specified, we use the simulation parameters described in this paragraph.
We assume $N_t=6$ transmit and $N_r=6$ receive antennas, and codewords of length $L=16$.
Both the transmit and receive arrays employ uniform linear arrays (ULAs) with half-wavelength spacing.
The transmit power, the communication receiver noise power, and the sensing receiver noise power are set to $\rho_t=\SI{30}{\decibel m}$, $\sigma_c^2=\SI{0}{\decibel m}$, and $\sigma_s^2=\SI{0}{\decibel m}$, respectively.
We assume the antenna \gls{xpd} is $\chi_{ant}=0.1$.
The target angle is $\theta_0=0^{\circ}$. Four clutter scatterers are located at angles $\phi_1=-80^{\circ}$, $\phi_2=-45^{\circ}$, $\phi_3=30^{\circ}$, and $\phi_4=60^{\circ}$.
The variance for the target and clutter attenuation is $\sigma_0^2=\sigma_q^2=\SI{-20}{\decibel m}$.
The target and clutter depolarization covariances are, respectively, given by \cite{jinOptimalPolarizationDesign2023,xiaoJointTransmitterReceiver2009}
\begin{equation*}
\begin{aligned}
\bm{\Sigma}_0&=
\begin{bmatrix}
    0.2 & 0.06\varepsilon & 0.05\varepsilon & 0.04\varepsilon \\
    0.06\varepsilon^* & 0.6 & 0.03\varepsilon & 0.03\varepsilon\\
    0.05\varepsilon^* & 0.03\varepsilon^* & 0.3 & 0.03\varepsilon \\
    0.04\varepsilon^* & 0.03\varepsilon^* & 0.03\varepsilon^* & 0.9
\end{bmatrix},
    \bm{\Sigma}_q=0.2\textbf{I}_4,
\end{aligned}
\end{equation*}
where $\varepsilon=1+j$.
The depolarization matrix for user $k$ is given by
\begin{equation*}
    \bm{\Phi}_{k,1}=\dots=\bm{\Phi}_{k,N_p}=
    \frac{1}{\sqrt{\chi_k+1}}\begin{bmatrix}
        e^{j\theta^{HH}_k} & \sqrt{\chi_k} e^{j\theta^{HV}_k} \\
         \sqrt{\chi_k}e^{j\theta^{VH}_k} & e^{j\theta^{VV}_k}
    \end{bmatrix},
\end{equation*}
where $\theta^{ij}_k$ is the phase shift from polarization $i$ to polarization $j$ and $\chi_k$ is the cross-polar power leakage of the channel for user $k$.
We set $\chi_k=0.1$ for all $k$.
The communication channel realizations are created using the Quasi
Deterministic Radio Channel Generator (QuaDRiGa), which provides realistic channel parameters drawn from measurement-based distributions \cite{jaeckel2014quadriga}.
The algorithm stopping thresholds are set to $\epsilon_1=10^{-5}$, $\epsilon_2=\epsilon_3=10^{-6}$, and $\epsilon_4=10^{-5}$.

For performance evaluation, we compare the proposed joint transmit/receive polarization optimization algorithm (``\textbf{Tx-Rx (Proposed)}'') with the following three baseline approaches. 
(i) Static polarization (``\textbf{Static}''): The transmit polarization vectors are predefined as $\textbf{p}_{t,n_t}=[1,0]^T$ for even-indexed antennas and $\textbf{p}_{t,n_t}=[0,1]^T$ for odd-indexed antennas. 
With this configuration, half of the antennas use vertical polarization while the other half use horizontal polarization, allowing the array to equally sample both polarizations.
The receive and user polarization vectors are configured in the same way.
(ii) Transmit-polarization-only optimization (``\textbf{Tx-only}''): The polarizations are optimized only at the transmitter, while the receive polarization vectors remain fixed as in the static polarization approach. 
(iii) Receive-polarization-only optimization (``\textbf{Rx-only}''): The polarizations are optimized only at the receiver, while the transmit polarization vectors remain fixed as in the static polarization approach. The same waveform optimization is applied for all methods.

\bb{
As a performance benchmark, we adapt our framework to a conventional dual-polarized \gls{mimo} system.
Specifically, the signal model in \eqref{eq:sensing_signal_mat} is modified as
\begin{equation}
\begin{aligned}
\textbf{Y}_{s,dual}&=\underbrace{\beta_0 {\textbf{A}_r(\theta_0)\bm{\Phi}_0\textbf{A}^T_t(\theta_0)}\textbf{X}_{dual}}_{\text{target response}} \\
&+
\underbrace{\displaystyle\sum_{q=1}^{N_c}\beta_q {\textbf{A}_r(\theta_q)\bm{\Phi}_q\textbf{A}^T_t(\theta_q)}\textbf{X}_{dual}}_{\text{clutter}}+\textbf{W}_{s,dual},
\end{aligned}
\end{equation}
where $\textbf{X}_{dual}=\textbf{F}_{dual}\textbf{S} \in \mathbb{C}^{N_{t,dual} \times L}$, $\textbf{F}_{dual} \in \mathbb{C}^{N_{t,dual} \times K}$ is the dual-polarized \gls{mimo} precoder and $N_{t,dual}$ is the total number of RF chains.
The corresponding \gls{mse} and target \gls{sinr} objectives can be obtained by defining the measurement matrix as $\textbf{M}_{dual}=\textbf{X}_{dual}^T \otimes \textbf{I}_{2N_r}$ and replacing $\textbf{M}$ in \eqref{eq:error_cov1} and \eqref{eq:metric:sinr} with $\textbf{M}_{dual}$.
Then, the dual-polarized \gls{mimo} precoder $\textbf{F}_{dual}$ can be optimized using the same \gls{sdr} algorithms for waveform design.
For the dual-polarized \gls{mimo} benchmark, we consider two antenna configurations: one utilizing the same number of RF chains as our proposed polarization-reconfigurable array ($N_{t,dual}=N_t$) but with half the array aperture, denoted as ``\textbf{dual-polarized MIMO (1$\times$)}'', and another configuration employing twice as many RF chains ($N_{t,dual}=2N_t$) while maintaining the same array aperture, denoted as ``\textbf{dual-polarized MIMO (2$\times$)}''.
}


\begin{figure}[!t]
\center{\includegraphics[width=.76\linewidth]{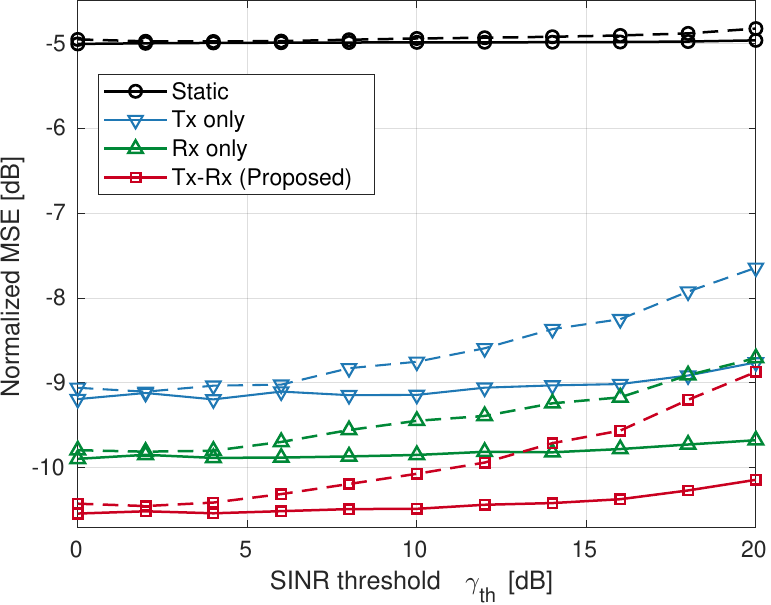}}
\caption{ \small 
Normalized \gls{mse} versus user \gls{sinr} threshold (Solid lines: $K=2$, dashed lines: $K=4$).
}
\label{fig:MSE_vs_sinr}\vspace{0.4 cm}
\end{figure}

\begin{figure}[!t]
\center{\includegraphics[width=.76\linewidth]{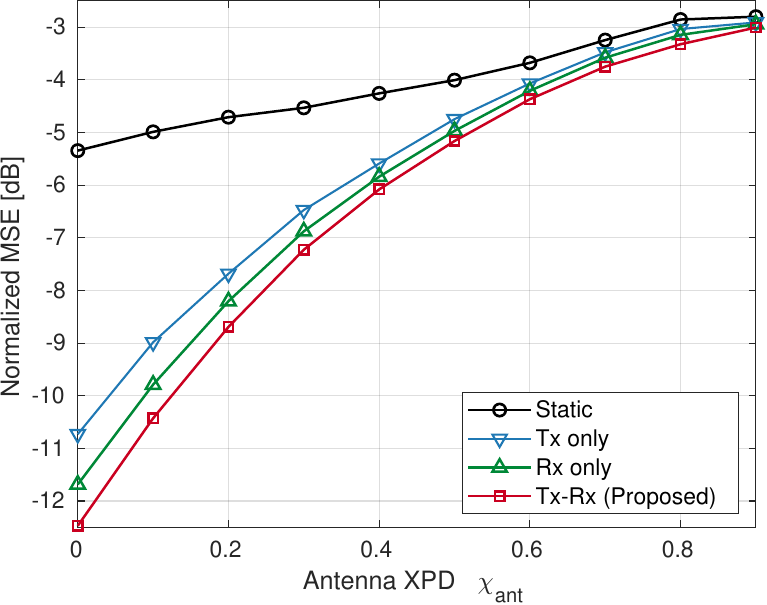}}
\caption{ \small 
Normalized \gls{mse} versus antenna \gls{xpd} $\chi_{ant}$ ($K=2$, $\gamma_{th}=\SI{10}{\decibel}$).
}
\label{fig:MSE_vs_xpd}\vspace{0.4 cm}
\end{figure}

\begin{figure}[!t]
\center{\includegraphics[width=.76\linewidth]{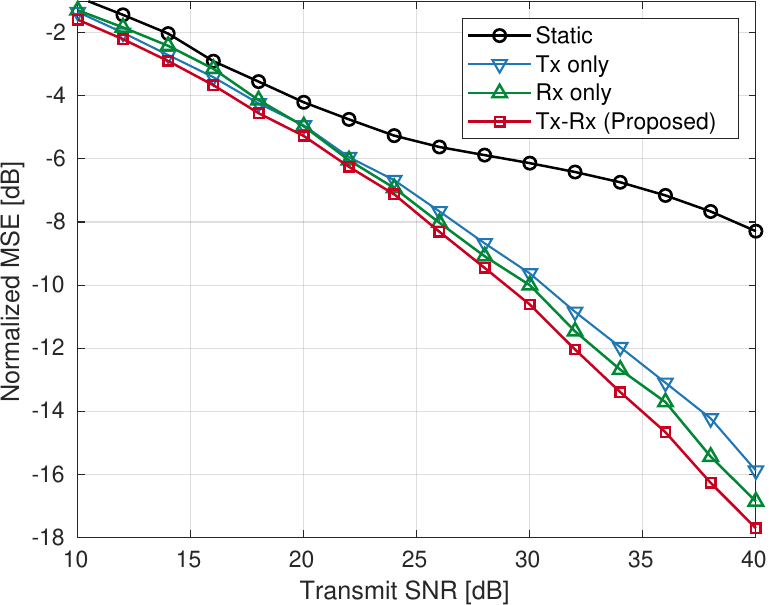}}
\caption{ \small 
Normalized \gls{mse} versus transmit SNR ($K=2$, $\gamma_{th}=10$dB). 
}
\label{fig:MSE_vs_SNR}\vspace{0.4 cm}
\end{figure}

\begin{figure}[!t]
\center{\includegraphics[width=.76\linewidth]{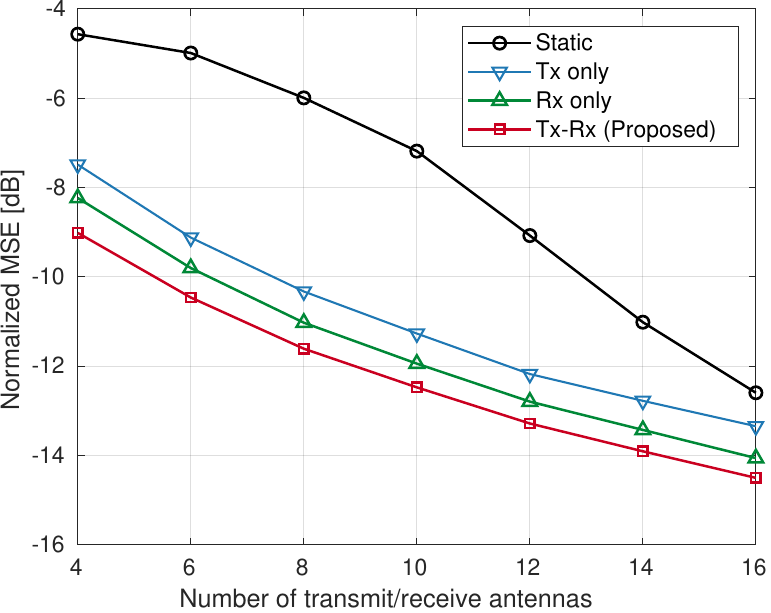}}
\caption{ \small 
Normalized \gls{mse} versus number of transmit/receive antennas ($K=2$, $\gamma_{th}=\SI{10}{\decibel}$).
}
\label{fig:MSE_vs_Nt}\vspace{0.4 cm}
\end{figure}

\begin{figure}[!t]
\center{\includegraphics[width=.76\linewidth]{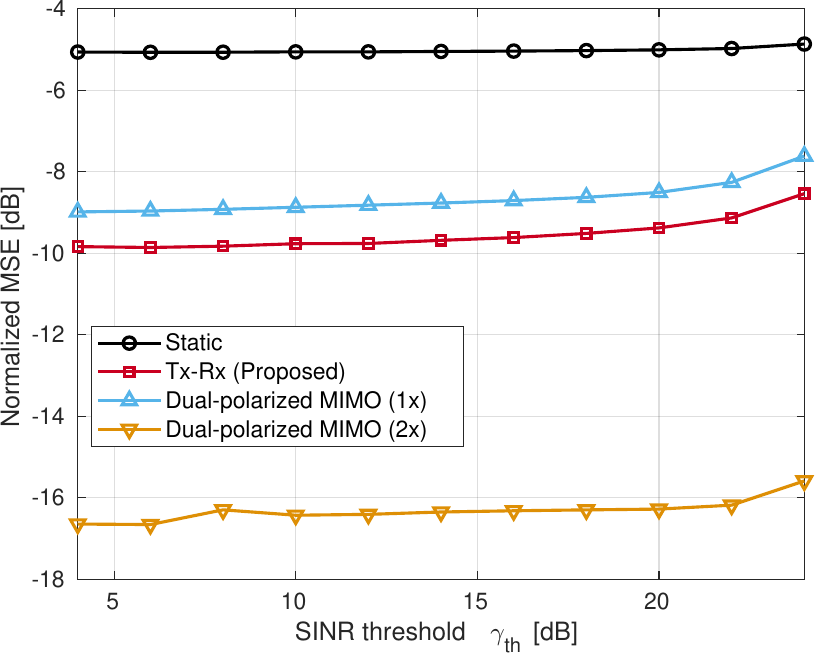}}
\caption{ \small 
Comparison of the static polarization, dual-polarized \gls{mimo} benchmarks, and the proposed Tx-Rx polarization optimization when $K=4$. 
}
\label{fig:MSE_dual_pol}
\end{figure}

\subsection{Normalized \Gls{mse} Performance}

In this section we evaluate the depolarization parameter estimation performance of the proposed algorithm. 
Fig. \ref{fig:MSE_vs_sinr} plots the normalized MSE of the depolarization estimate for the proposed and baseline approaches for varying user \gls{sinr} thresholds under scenarios with  $K=2,~4$ users.
The \gls{nmse} is computed as
\begin{equation}
    \text{Normalized MSE}=10\log_{10}\frac{\operatorname{Tr}(\textbf{Z}_e)}{\operatorname{Tr}(\bm{\Sigma}_0)}.
\end{equation}
As expected, the algorithms that optimize the polarization at one or both ends of the link significantly outperform the case with static polarization.
The \gls{nmse} increases with the \gls{sinr} threshold for the cases with optimized polarization, indicating the inherent trade-off between sensing accuracy and communication performance.
Moreover, the \gls{nmse} is higher for $K=4$ than $K=2$ since the presence of more users reduces the flexibility of the sensing precoder.
Interestingly, optimizing only the Rx polarization leads to lower MSE than for the Tx-only case.
This is primarily because the target depolarization gains for the vertically polarized receiver antennas, i.e., $HV$ and $VV$, are larger than $HH$ and $VH$.


Fig. \ref{fig:MSE_vs_xpd} plots the \gls{nmse} of the depolarization estimate versus the antenna \glspl{xpd}.
When $\chi_{ant}=0$, the antennas can perfectly distinguish between co-polarized and cross-polarized signals, whereas when $\chi_{ant}=1$ antennas lose all discrimination capability. Consequently, the \gls{nmse} performance deteriorates as $\chi_{ant} \rightarrow 1$ due to increased power leakage between the H and V polarization channels.
Such leakage negatively affects both sensing and communication performance. 
The performance gap between the joint Tx-Rx optimization and the static polarization case narrows as \gls{xpd} increases, eventually vanishing at $\chi_{ant}=1$.
This result is intuitive since the polarized channels become increasingly correlated as $\chi_{ant} \rightarrow 1$, reducing the number of polarization-based DoFs.
For the Tx-Rx optimization, the performance gap between $\chi_{ant}=0$ and $\chi_{ant}=1$ exceeds $\SI{5}{\decibel}$, suggesting its significant impact on polarimetric sensing quality.

Fig. \ref{fig:MSE_vs_SNR} illustrates the \gls{nmse} performance versus the transmit \gls{snr}.
As expected, the \gls{nmse} performance of all approaches improves as the \gls{snr} increases.
However, the static polarization scheme experiences diminishing returns with increasing \gls{snr}, eventually saturating at approximately $\SI{-8}{\decibel}$.
In contrast, the algorithms that optimally configure the polarization continue to improve with higher transmit \gls{snr}, with the performance gap growing to nearly \SI{10}{\decibel} at a transmit \gls{snr} of $\SI{40}{\decibel}$.

Fig. \ref{fig:MSE_vs_Nt} plots the normalized \gls{nmse} for different numbers of transmit and receive antennas when $N_t=N_r$.
In line with the previous results, the joint Tx-Rx optimization consistently outperforms the Rx-only and Tx-only approaches by about \SI{0.5}{\decibel} and \SI{1}{\decibel}, respectively. In this case, the performance gap between the static and reconfigurable polarization algorithms decreases with the number of antennas.
The performance gain is about $\SI{3}{\decibel}$ for $N_t=4$ and grows to $\SI{4}{\decibel}$ when $N_t=8$, after which it decreases.
This suggests that polarization optimization is critical to improve the polarimetric sensing performance, particularly with smaller arrays.

\bb{
Fig. \ref{fig:MSE_dual_pol} compares the \gls{nmse} for the static polarization case, the dual-polarized \gls{mimo} benchmarks, and the proposed Tx-Rx polarization optimization as a function of the user \gls{sinr} threshold when $K=4$.
For this evaluation,  we configured three clutter objects at angles $\phi_1=-70^{\circ}$, $\phi_2=-15^{\circ}$, and $\phi_3=65^{\circ}$.
As expected, the 2$\times$ dual-polarized \gls{mimo} benchmark yields the lowest \gls{nmse} among all schemes since it employs twice as many RF chains as the proposed polarization-reconfigurable array.
In contrast, the proposed Tx-Rx optimization outperforms the 1$\times$ dual-polarized \gls{mimo} benchmark since, with the same number of RF chains, it maintains the full array aperture to maximize the available spatial DoFs. 
}


\begin{figure}[!t]
\center{\includegraphics[width=.8\linewidth]{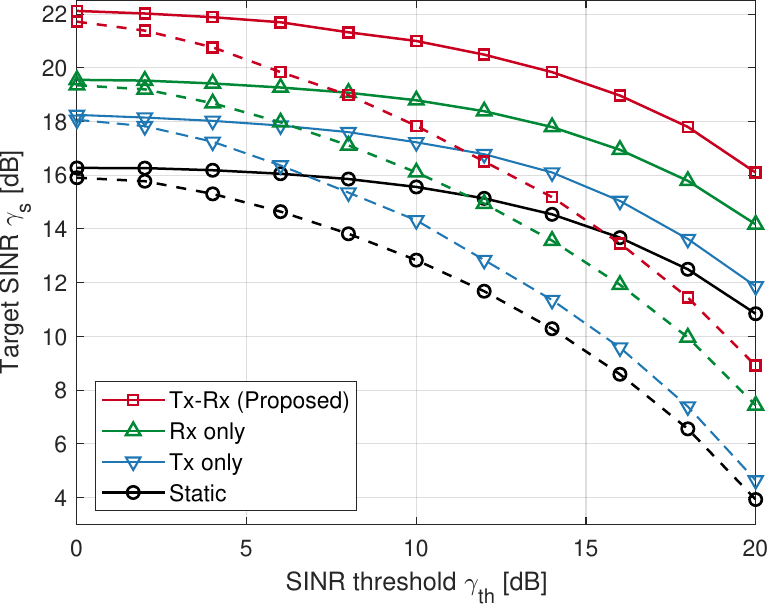}}
\caption{ \small 
Target \gls{sinr} versus user \gls{sinr} threshold
 (Solid lines: $K=2$, dashed lines: $K=4$).
}
\label{fig:SINR_vs_SINR}\vspace{0.4 cm}
\end{figure}

\begin{figure}[!t]
\center{\includegraphics[width=.8\linewidth]{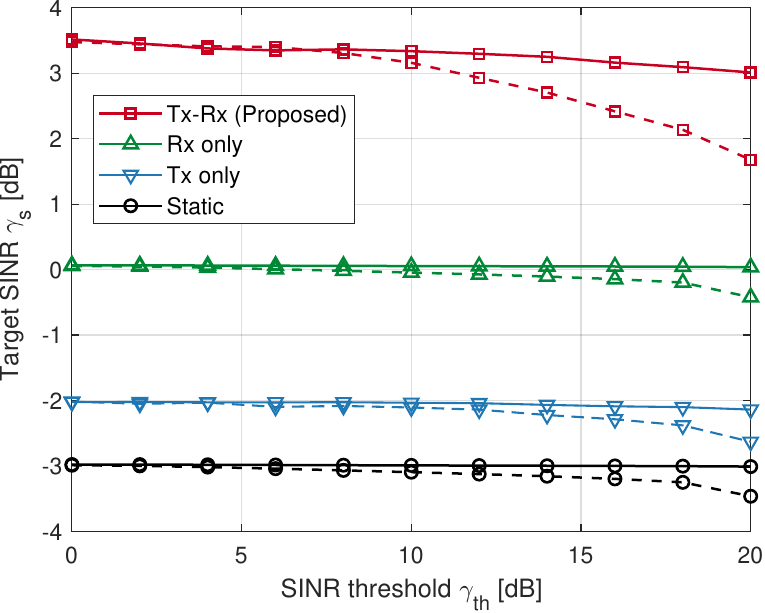}}
\caption{ \small 
Target \gls{sinr} versus user \gls{sinr} threshold under a clutter-rich scenario (Solid lines: $K=2$, dashed lines: $K=4$).
}
\label{fig:SINR_vs_SINR_clutter}\vspace{0.4 cm}
\end{figure}

\subsection{\Gls{sinr} Performance}
Here we evaluate the target \gls{sinr} performance of the proposed approach.
The target depolarization covariance was set to
\begin{equation}
    \bm{\Sigma}_0=
\begin{bmatrix}
    0.1 & 0.06\varepsilon & 0.05\varepsilon & 0.04\varepsilon \\
    0.06\varepsilon^* & 0.3 & 0.03\varepsilon & 0.03\varepsilon\\
    0.05\varepsilon^* & 0.03\varepsilon^* & 0.1 & 0.03\varepsilon \\
    0.04\varepsilon^* & 0.03\varepsilon^* & 0.03\varepsilon^* & 0.9
\end{bmatrix},
\end{equation}
which represents a target with a strong vertical co-polar ($VV$) return (e.g., building), allowing clear separation from clutter.
The transmit, receive, and user polarizations for the static polarization case are all set to horizontal polarization.
Fig. \ref{fig:SINR_vs_SINR} shows the target \gls{sinr} for different user \gls{sinr} thresholds.
The proposed Tx-Rx joint optimization yields the highest target SINRs due to its ability to exploit all available polarization DoFs, allowing better clutter suppression and enhanced target response. 
In particular, the joint Tx-Rx optimization achieves a performance gain of approximately $\SI{5.5}{\decibel}$ - $\SI{6}{\decibel}$ over the static polarization case, although this advantage decreases slightly as the \gls{qos} threshold increases. 
Consistent with the findings for \gls{mse}, the target \gls{sinr} decreases as the user \gls{sinr} threshold grows or the number of users increases due to the sensing-communication trade-off. 
Similar to the \gls{mse} minimization scenario, the Rx-only optimization generally achieves a target \gls{sinr} gain of $\SI{1.5}{\decibel}$ - $\SI{2}{\decibel}$ over the Tx-only case, due to the larger gain in the $VV$ channel.

Fig. \ref{fig:SINR_vs_SINR_clutter} shows the target \gls{sinr} for different user \gls{sinr} thresholds in a challenging scenario where four clutter objects are more closely located around the target, at angles $\phi_1=-15^{\circ}$, $\phi_2=-5^{\circ}$, $\phi_3=5^{\circ}$, and $\phi_4=15^{\circ}$.
We observe that the \gls{sinr} values are lower than those in Fig. \ref{fig:SINR_vs_SINR}, reflecting the increased difficulty in discriminating the target from closely situated clutter.
Notably, the performance gap between the proposed Tx-Rx joint polarization optimization and the Tx-only or Rx-only approaches is significantly larger in this severe clutter scenario, reaching about a $\SI{6.5}{\decibel}$ gain.
These observations suggest that joint optimization of the transmit and receive polarizations is crucial to fully exploit the potential of polarization reconfigurability in heavily cluttered environments.

\begin{figure}[!t]
\center{\includegraphics[width=.8\linewidth]{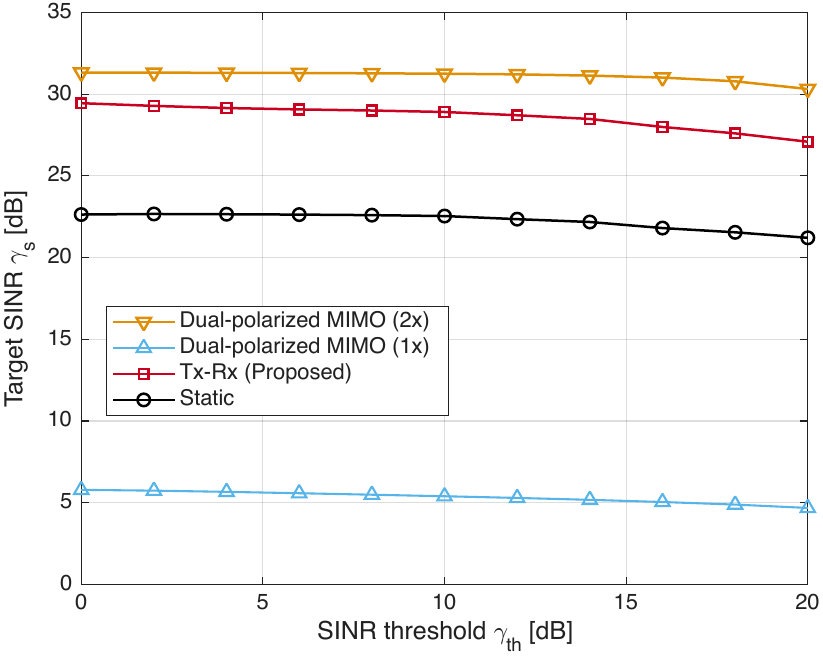}}
\caption{ \small 
Target \gls{sinr} versus user \gls{sinr} threshold under the same clutter setting as in Fig. \ref{fig:SINR_vs_SINR} for $N_t=16$.
}
\label{fig:SINR_vs_SINR_16Tx}\vspace{0.4 cm}
\end{figure}

\bb{
Fig. \ref{fig:SINR_vs_SINR_16Tx} compares the target \gls{sinr} for the static, dual-polarized \gls{mimo} baselines and the proposed polarization reconfiguration scheme under the same clutter configuration as in Fig. \ref{fig:SINR_vs_SINR}.
The \gls{bs} transmit and receive arrays are configured with $N_t=16$ and $N_r=6$ antennas respectively, which implies 32 total transmit antennas for the dual-polarized \gls{mimo} 2$\times$ benchmark, representing a typical massive \gls{mimo} scenario.
The number of communication users is set to $K=4$.
The overall target \gls{sinr} level increases compared to the scenario in Fig. \ref{fig:SINR_vs_SINR} due to the increased number of transmit antennas, i.e., higher array gains.
The polarization-reconfiguration scheme shows comparable performance to the dual-polarized \gls{mimo} 2$\times$ benchmark, achieving more than a $\SI{5}{\decibel}$ improvement in target \gls{sinr} over the static polarization benchmark.
In contrast, the dual-polarized \gls{mimo} 1$\times$ baseline performs notably worse, primarily due to reduced spatial resolution resulting from its smaller array aperture.
}

\begin{figure}[!t]
\center{\includegraphics[width=.8\linewidth]{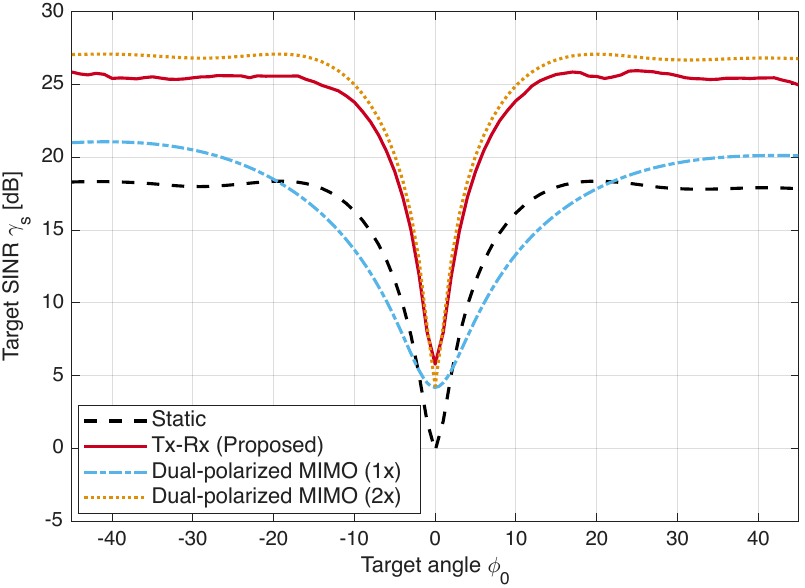}}
\caption{ \small 
Target \gls{sinr} with varying target angles under a scenario where a single clutter object is placed at an angle $0^{\circ}$. 
We set $K=2$ and $\gamma_{th}=\SI{10}{\decibel}$.
}
\label{fig:SINR_vs_angle}
\end{figure}

\begin{figure}[!t]
\center{\includegraphics[width=.8\linewidth]{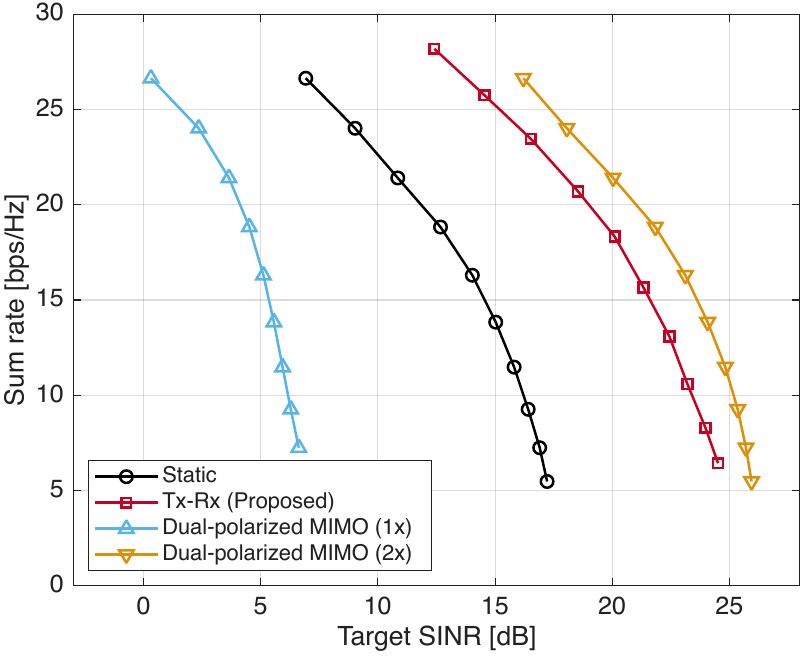}}
\caption{ \small 
Target \gls{sinr} versus sum rate for the static, dual-polarized \gls{mimo} and proposed Tx-Rx polarization optimization schemes when $K=4$. 
We use the same cluttering setting in Fig. \ref{fig:SINR_vs_SINR}.
}
\label{fig:capcity_vs_SINR}\vspace{0.4 cm}
\end{figure}

\bb{
Fig. \ref{fig:SINR_vs_angle} shows the target \gls{sinr} as a function of the target angle between $-45^{\circ}$ to $45^{\circ}$ for static polarization, dual-polarized \gls{mimo}, and the proposed Tx-Rx optimization. A single clutter source is fixed at an angle of $0^{\circ}$.
Notably, all schemes experience a deep \gls{sinr} drop when the target angle approaches the clutter direction.
Conversely, when the target and clutter are well-separated in the angular domain, both the proposed polarization-reconfigurable design and the dual-polarized \gls{mimo} (2$\times$) benchmark achieve significantly higher \gls{sinr} levels than the static polarization baseline.
Furthermore, the dual-polarized \gls{mimo} (1$\times$) benchmark yields a wider \gls{sinr} null compared to the proposed reconfigurable Tx-Rx polarization approach, reflecting its reduced angular resolution and thus inferior target discrimination capability.
}

\bb{
Fig. \ref{fig:capcity_vs_SINR} plots the trade-off between the communication sum rate and target \gls{sinr} for the static polarization case, the dual-polarized \gls{mimo} benchmarks, and the proposed polarization-reconfigurable approaches when $K=4$.
As can be seen, in all cases, target \gls{sinr} values decrease as the sum rate increases due to the sensing-communication trade-off.
The proposed Tx-Rx optimization substantially improves the Pareto frontier compared to the static polarization case.
For example, at a sum rate of $10\ \text{bps/Hz}$, the optimized Tx-Rx polarization configuration achieves an SINR gain exceeding $\SI{10}{\decibel}$.
Moreover, the proposed Tx-Rx optimized polarization approaches the performance of the dual-polarized \gls{mimo} 2$\times$ benchmark, demonstrating that polarization reconfigurability can effectively exploit polarization-domain DoFs without requiring twice the number of RF chains.
In line with the previous results, the dual-polarized \gls{mimo} 1$\times$ baseline shows the worst performance among all schemes, which results from its limited angular resolution.
}



\section{Conclusion}
In this paper, we explored waveform and polarization optimization for \gls{ipsac} systems with polarization-reconfigurable arrays.
We formulated optimization problems with the aim of minimizing the \gls{mse} of the target depolarization parameter or maximizing the target \gls{sinr}, subject to the communication SINR requirements and power budget.
To tackle these challenging nonconvex problems, we developed effective algorithms leveraging \gls{sdr}, \gls{mm}, and Dinkelbach's transform.
Extensive simulation results demonstrate that the proposed joint waveform and polarization optimization approach achieves significant performance improvements compared with systems that do not exploit reconfigurable polarizations for the transmit or receive antennas. 
In particular, the proposed algorithm can substantially reduce the \gls{mse} of the depolarization parameter estimate by several dB compared to static polarization methods. Moreover, substantial enhancements in target SINR were observed, indicating the robustness of the proposed method in challenging and clutter-rich scenarios.
\bb{Moreover, comparisons with dual-polarized MIMO benchmarks show that the proposed polarization-reconfiguration scheme achieves target \gls{sinr} performance close to a dual-polarized MIMO array employing twice as many RF chains, while maintaining the same number of RF chains as a single-polarized system.}

\appendices

\section{Proof of Lemma \ref{theorem:majorization}}\label{sec:appendix_A}
Let $\textbf{E}_i=\bar{\textbf{X}}^H\bar{\textbf{P}}_i^T\bm{\Omega}\bar{\textbf{P}}_i\bar{\textbf{X}}+\sigma_s^2\textbf{I}_{LN_r}$ and $\textbf{B}_i=\bar{\textbf{X}}^H\bar{\textbf{P}}_i^T\textbf{A}_0{\bm{\Sigma}}_0$.
The objective can be rewritten as $-\operatorname{Tr}(\textbf{B}^H\textbf{E}^{-1}\textbf{B})$, which is jointly concave in \textbf{B} and \textbf{E}.
Hence, using a first-order Taylor expansion, the objective can be majorized at point $(\textbf{B}_{i},\textbf{E}_{i})$ as \cite{sunMajorizationMinimizationAlgorithmsSignal2017a} \vspace{-3mm}
\begin{equation}
\begin{aligned}
    &-\operatorname{Tr}(\textbf{B}^H\textbf{E}^{-1}\textbf{B}) \\
    &\leq \operatorname{Tr}(\textbf{E}^{-1}_i\textbf{B}_{i}\textbf{B}_{i}^H\textbf{E}_{i}^{-1}\textbf{E}) - 2\operatorname{Re}(\operatorname{Tr}(\textbf{B}_{i}^H\textbf{E}_{i}^{-1}\textbf{B}))+const \\
    &= \operatorname{Tr}(\textbf{B}_{i}^H\textbf{E}_{i}^{-1}\bar{\textbf{X}}^H\bar{\textbf{P}}^T\bm{\Omega}\bar{\textbf{P}}\bar{\textbf{X}}\textbf{E}^{-1}_i\textbf{B}_{i}) \\
    &\quad - 2\operatorname{Re}(\operatorname{Tr}(\textbf{B}_{i}^H\textbf{E}_{i}^{-1}\bar{\textbf{X}}^H\bar{\textbf{P}}^T\textbf{A}_0\bm{\Sigma}_0))+const.
\end{aligned}    
    \end{equation}
     Let $\textbf{D}=\bar{\textbf{P}}\bar{\textbf{X}}\textbf{E}^{-1}\textbf{B}$.
    Then, $\operatorname{Tr}(\textbf{D}^H\bm{\Omega}\textbf{D})$ can be majorized  at point $\textbf{D}_{i}$ using a first-order Taylor expansion as \cite[Equation~13]{sunMajorizationMinimizationAlgorithmsSignal2017a}
    \begin{equation}
        \begin{aligned}
            &\operatorname{Tr}(\textbf{D}^H\bm{\Omega}\textbf{D}) \\
            &\leq 
            \lambda_{\bm{\Omega}}\operatorname{Tr}(\textbf{D}_{i}^H\textbf{D}_{i})+2\operatorname{Re}\left(\operatorname{Tr}\left(\textbf{D}^H(\bm{\Omega}-\lambda_{\bm{\Omega}}\textbf{I}_{4N_tN_r})\textbf{D}_{i}\right)\right)\\
            & \quad +\operatorname{Tr}(\textbf{D}_{i}^H(\lambda_{\bm{\Omega}}\textbf{I}_{4N_tN_r}-\bm{\Omega})\textbf{D}_{i}) \\            
            &= 
            2\operatorname{Re}(\operatorname{Tr}(\bar{\textbf{P}}^T(\bm{\Omega}-\lambda_{\bm{\Omega}}\textbf{I}_{4N_tNr})\textbf{D}_{i}\textbf{B}_{i}^H\textbf{E}_{i}^{-1}\bar{\textbf{X}}^H)) + const.
        \end{aligned}
    \end{equation}
Notice $\operatorname{Tr}(\textbf{D}_{i}^H\textbf{D}_{i})$ is a constant because 
\begin{equation*}   
\begin{aligned}    \textbf{D}^H_i\textbf{D}_{i}&=\textbf{B}_{i}^H(\textbf{E}_{i}^{-1})^H\bar{\textbf{X}}^H{\bar{\textbf{P}}_i^T\bar{\textbf{P}}_i}_{}\bar{\textbf{X}}\textbf{E}_{i}^{-1}\textbf{B}_{i}\\
&=\textbf{B}_{i}^H(\textbf{E}_{i}^{-1})^H\bar{\textbf{X}}^H\bar{\textbf{X}}\textbf{E}_{i}^{-1}\textbf{B}_{i},
\end{aligned}
\end{equation*}
where the second equality comes from the fact that $\bar{\textbf{P}}_i^T\bar{\textbf{P}}_i=\textbf{I}_{N_tN_r}$.
Combining the above results yields 
\begin{equation}
    \begin{aligned}
        &-\operatorname{Tr}(\textbf{B}^H\textbf{E}^{-1}\textbf{B})
        \\
        &\leq \quad 2\operatorname{Re}\big(\operatorname{Tr}\big(\bar{\textbf{P}}^T(\bm{\Omega}-\lambda_{\bm{\Omega}}\textbf{I}_{4N_tN_r})\textbf{D}_{i}\textbf{B}_{i}^H\textbf{E}_{i}^{-1}\bar{\textbf{X}}^H\big)\big) \\
        & \quad - 2\operatorname{Re}\big(\operatorname{Tr}\big(\textbf{B}_{i}^H\textbf{E}_{i}^{-1}\bar{\textbf{X}}^H\bar{\textbf{P}}^T\textbf{A}_0\bm{\Sigma}_0\big)\big)+const\\
        &= \quad \operatorname{Tr}(\bar{\textbf{P}}^T\bm{\Lambda}_i )+const,
    \end{aligned}
\end{equation}
where $\bm{\Lambda}_i = 2\operatorname{Re}\big([\left(\bm{\Omega}-\lambda_{\bm{\Omega}}\textbf{I}_{4N_tN_r}\right)\textbf{D}_{i}
        -\textbf{A}_0\bm{\Sigma}_0]\textbf{B}_{i}^H\textbf{E}_{i}^{-1}\bar{\textbf{X}}^H\big)$
and the equality follows from the fact that $\bar{\textbf{P}}$ is a real-valued matrix.

\section{Proof of Lemma \ref{lemma:mse_majorized}}\label{appendix:proof_obj_pt}
The majorizer can be represented as
\begin{equation}\label{eq:mse_majorized}
    \begin{aligned}        \operatorname{Tr}\big(\bar{\textbf{P}}^T\bm{\Lambda}_i\big)=\operatorname{Tr}\big(\bar{\textbf{P}}_t^T\bar{\textbf{P}}_r^T\bm{\Lambda}_i\big),
    \end{aligned}
\end{equation}
where $\bar{\textbf{P}}_t=\textbf{P}_t \otimes \textbf{I}_{N_r}$ and $\bar{\textbf{P}}_r=\textbf{I}_{N_t} \otimes \textbf{P}_{r}$.
By leveraging the properties of the Kronecker product, the vectorization of matrix $\bar{\textbf{P}}$ can be represented as
\begin{equation}
    \begin{aligned}
        \operatorname{vec}(\textbf{P}_t \otimes \textbf{I}_{N_r})&= {\bar{\textbf{I}}_t}\operatorname{vec}(\textbf{P}_t),\quad 
        \operatorname{vec}(\textbf{I}_{N_t} \otimes \textbf{P}_r)= \bar{\textbf{I}}_r\operatorname{vec}(\textbf{P}_r),
    \end{aligned}
\end{equation}
where 
\begin{equation}
\begin{aligned}
    \bar{\textbf{I}}_t&=\textbf{I}_{N_t}\otimes(\textbf{K}_{N_r,2N_t} \otimes \textbf{I}_{N_r})(\textbf{I}_{2N_t} \otimes \operatorname{vec}(\textbf{I}_{N_r})),\\
    \bar{\textbf{I}}_r&=(\textbf{I}_{N_t} \otimes \textbf{K}_{N_r,2N_t} )(\operatorname{vec}(\textbf{I}_{N_t}) \otimes \textbf{I}_{N_r})\otimes \textbf{I}_{2N_r},
\end{aligned}    
\end{equation}
and $\textbf{K}_{N_r,2N_t}\in\{0,1\}^{2N_tN_r\times 2N_tN_r}$ is the commutation matrix \cite{henderson1981vec}.
The block-diagonal structure of $\textbf{P}_t$ and $\textbf{P}_r$ allows them to be rewritten as
\begin{equation}
\begin{aligned}
    \operatorname{vec}(\textbf{P}_t)&=[\textbf{e}_{N_t,1}^T\otimes \textbf{p}_{t,1}^T, \textbf{e}_{N_t,2}^T\otimes \textbf{p}_{t,2}^T,\ldots,\textbf{e}_{N_t,N_t}^T\otimes \textbf{p}_{t,N_t}^T]^T\\
    &=\bm{\Theta}_t
    \textbf{p}_t,\\
    \operatorname{vec}(\textbf{P}_r)&=[\textbf{e}_{N_r,1}^T\otimes \textbf{p}_{r,1}^T, \textbf{e}_{N_r,2}^T\otimes \textbf{p}_{r,2}^T,\ldots,\textbf{e}_{N_r,N_r}^T\otimes \textbf{p}_{r,N_r}^T]^T\\
    &=\bm{\Theta}_r
    \textbf{p}_r,
\end{aligned}    
\end{equation}
where 
$\bm{\Theta}_t=\text{blkdiag}(\textbf{e}_{N_t,1},\textbf{e}_{N_t,2},\dots,\textbf{e}_{N_t,N_t})\otimes \textbf{I}_2$, $ \bm{\Theta}_r=\text{blkdiag}(\textbf{e}_{N_r,1},\textbf{e}_{N_r,2},\dots,\textbf{e}_{N_r,N_r})\otimes \textbf{I}_2$, and $\textbf{e}_N,i$ is the $i$th column of $\textbf{I}_N$.
This allows Eq. \eqref{eq:mse_majorized} to be rewritten as
\begin{equation}
    \begin{aligned}        \operatorname{Tr}\big(\bar{\textbf{P}}^T\bm{\Lambda}_i\big)&=\operatorname{Tr}\big(\bar{\textbf{P}}_t^T\bar{\textbf{P}}_r^T\bm{\Lambda}_i\big)=\operatorname{vec}^T\big(\bar{\textbf{P}}_t\big)\operatorname{vec}\big(\bar{\textbf{P}}_r^T\bm{\Lambda}_i\big) \\
    &=\textbf{p}^T_t\underbrace{\bm{\Theta}_t^T\bar{\textbf{I}}_t^T\operatorname{vec}(\bar{\textbf{P}}_r^T\bm{\Lambda}_i)}_{\textbf{d}^{(i)}}=\textbf{p}_t^T\textbf{d}^{(i)}.
    \end{aligned}
\end{equation}
Similarly, we can write
\begin{equation}
    \begin{aligned}        \operatorname{Tr}\big(\bar{\textbf{P}}^T\bm{\Lambda}_i\big)&=\operatorname{Tr}\big(\bar{\textbf{P}}_r^T\bar{\textbf{P}}_t^T\bm{\Lambda}_i\big)=\operatorname{vec}^T\big(\bar{\textbf{P}}_r\big)\operatorname{vec}\big(\bar{\textbf{P}}_t^T\bm{\Lambda}_i\big) \\
    &=\textbf{p}^T_r{\bm{\Theta}_r^T\bar{\textbf{I}}_r^T\operatorname{vec}\big(\bar{\textbf{P}}_t^T\bm{\Lambda}_i\big)}_{}=\textbf{p}_r^T\textbf{q}^{(i)}
    \end{aligned}
\end{equation}
where $\textbf{q}^{(i)}=\bm{\Theta}_r^T\bar{\textbf{I}}_r^T\operatorname{vec}\big(\bar{\textbf{P}}_t^T\bm{\Lambda}_i\big)$.


\section{Proof of Lemma \ref{lemma:sinr_constraint}}\label{appendix:proof_sinr_const}

The \gls{sinr} for user $k$ is given by 
\begin{equation}
\begin{aligned}
    \gamma_k &= \frac{|\tilde{\textbf{h}}_k^H\textbf{P}_t\textbf{f}_k|^2}{\sum_{\substack{k' \neq k}}^K|\tilde{\textbf{h}}_k^H\textbf{P}_t\textbf{f}_k'|^2+\sigma_c^2}\\ 
    &=\frac{\operatorname{vec}^T(\textbf{P}_t)\big(\textbf{f}^*_{k}\textbf{f}_{k}^T\otimes \tilde{\textbf{h}}_k\tilde{\textbf{h}}_k^H\big)\operatorname{vec}(\textbf{P}_t)}{\sum_{\substack{k' \neq k}}^K \operatorname{vec}^T(\textbf{P}_t)\big(\textbf{f}^*_{k'}\textbf{f}_{k'}^T\otimes \tilde{\textbf{h}}_k\tilde{\textbf{h}}_k^H\big)\operatorname{vec}(\textbf{P}_t)+\sigma_c^2},
\end{aligned}
    \end{equation}
where $\tilde{\textbf{h}}_k^H\triangleq \textbf{p}^T_u\textbf{H}_{up,k}$ and the second equality holds since    $\tilde{\textbf{h}}_k^H\textbf{P}_t\textbf{f}_k=(\textbf{f}_k^T\otimes \tilde{\textbf{h}}_k^H)\operatorname{vec}(\textbf{P}_t)$.
The \gls{sinr} constraint can be converted to
\begin{equation}\label{ineq:sinr_Pt}
    \begin{aligned}         
    \operatorname{vec}^T(\textbf{P}_t)\bm{\Psi}_k\operatorname{vec}(\textbf{P}_t)+\gamma_{th}\sigma_c^2\leq 0,
    \end{aligned}
\end{equation}
where $ \bm{\Psi}_k= \big(\gamma_{th}\sum\nolimits_{\substack{ k' \neq k}}^K\textbf{f}^*_{k'}\textbf{f}_{k'}^T -\textbf{f}^*_{k}\textbf{f}_{k}^T\big) \otimes \tilde{\textbf{h}}_k\tilde{\textbf{h}}_k^H.$
This allows the quadratic term in \eqref{ineq:sinr_Pt} to be rewritten as
\begin{equation}
   \operatorname{vec}^T(\textbf{P}_t)\bm{\Psi}_k\operatorname{vec}(\textbf{P}_t) 
   = \textbf{p}_t^T {\bm{\Theta}_t^T{\bm{\Psi}}_k\bm{\Theta}_t}_{} \textbf{p}_t
   = \textbf{p}_t^T \bar{\bm{\Psi}}_k \textbf{p}_t.
\end{equation}
where $\bar{\bm{\Psi}}_k=\bm{\Theta}_t^T{\bm{\Psi}}_k\bm{\Theta}_t$.
Thus, the \gls{sinr} constraint can be expressed as $\textbf{p}_t^T \bar{\bm{\Psi}}_k \textbf{p}_t+\gamma_{th}\sigma^2_c \leq 0$.

\section{Derivation of \eqref{prob:sinr_lp}}\label{appendix:proof_sinr_majorizer}
Similar to the derivation in Appendix \ref{appendix:proof_obj_pt}, 
$\textbf{g}^{(i)}$ can be obtained as 
\begin{equation}
\begin{aligned}
    &\operatorname{Tr}\big(\bar{\textbf{P}}^T(\bm{\Xi}_i-\lambda_{\bm{\Xi}_i}\textbf{I})\bar{\textbf{P}}_i\bar{\textbf{R}}_x\big) \\&= \operatorname{Tr}\big(\bar{\textbf{P}}_t^T\bar{\textbf{P}}_r^T(\bm{\Xi}_i-\lambda_{\bm{\Xi}_i}\textbf{I})\bar{\textbf{P}}_i\bar{\textbf{R}}_x\big)\\ &=\textbf{p}^T_t\underbrace{\bm{\Theta}_t^T\bar{\textbf{I}}_t^T\operatorname{vec}\big(\bar{\textbf{P}}_r^T(\bm{\Xi}_i-\lambda_{\bm{\Xi}_i}\textbf{I})\bar{\textbf{P}}_i\bar{\textbf{R}}_x\big)}_{\textbf{g}^{(i)}}=\textbf{p}_t^T\textbf{g}^{(i)}.
\end{aligned}    
\end{equation}
Similarly, we have $\operatorname{Tr}\big(\bar{\textbf{P}}^T(\bm{\Xi}_i-\lambda_{\bm{\Xi}_i}\textbf{I})\bar{\textbf{P}}_i\bar{\textbf{R}}_x\big)=\textbf{p}_r^T\textbf{v}^{(i)}$ where $\textbf{v}^{(i)}=\bm{\Theta}_r^T\bar{\textbf{I}}_r^T\operatorname{vec}\big(\bar{\textbf{P}}_t^T(\bm{\Xi}_i-\lambda_{\bm{\Xi}_i}\textbf{I})\bar{\textbf{P}}_i\bar{\textbf{R}}_x\big)$.

\bibliographystyle{IEEEtran}
\bibliography{IEEEabrv,references}

\end{document}